\documentclass[twocolumn,english,aps,prl,superscriptaddress,floats,nobibnotes]{revtex4-1}
\usepackage{amsmath}
\usepackage{amssymb}
\usepackage{graphicx}
\usepackage{xcolor}
\usepackage[normalem]{ulem}

\begin{document}

\title{Evidence of Rotational Fr\"ohlich Coupling in Polaronic Trions}

\author{Maxim Trushin}
\affiliation{Centre for Advanced 2D Materials, National University of Singapore, Singapore 117546} 

\author{Soumya Sarkar}
\affiliation{NUSNNI NanoCore, National University of Singapore, Singapore 117411}

\author{Sinu Mathew$^\dagger$}

\affiliation{NUSNNI NanoCore, National University of Singapore, Singapore 117411}

\author{Sreetosh Goswami}
\affiliation{NUSNNI NanoCore, National University of Singapore, Singapore 117411}

\author{Prasana Sahoo}
\affiliation{Department of Materials Science and Metallurgy, University of Cambridge, Cambridge, UK CB30FS} 

\author{Yan Wang}
\affiliation{Department of Materials Science and Metallurgy, University of Cambridge, Cambridge, UK CB30FS}

\author{Jieun Yang}
\affiliation{Department of Materials Science and Metallurgy, University of Cambridge, Cambridge, UK CB30FS} 

\author{Weiwei Li}
\affiliation{Department of Materials Science and Metallurgy, University of Cambridge, Cambridge, UK CB30FS} 

\author{Judith L. MacManus-Driscoll}
\affiliation{Department of Materials Science and Metallurgy, University of Cambridge, Cambridge, UK CB30FS} 

\author{Manish Chhowalla}
\affiliation{Department of Materials Science and Metallurgy, University of Cambridge, Cambridge, UK CB30FS} 

\author{Shaffique Adam$^*$}
\affiliation{Centre for Advanced 2D Materials, National University of Singapore, Singapore 117546}
\affiliation{Department of Physics, National University of Singapore, Singapore, 117551}
\affiliation{Yale-NUS College, Singapore 138527}

\author{T. Venkatesan$^*$}
\affiliation{NUSNNI NanoCore, National University of Singapore, Singapore 117411}
\affiliation{Department of Physics, National University of Singapore, Singapore, 117551}
\affiliation{Department of Electrical and Computer Engineering and Materials Science and Engineering, National University of Singapore, Singapore 117583}

\date{\today}

\begin{abstract}

Electrons commonly couple through  Fr\"ohlich interactions with longitudinal optical phonons to form polarons.
However, trions possess a finite angular momentum and should therefore couple instead to rotational optical phonons.
This creates a polaronic trion whose binding energy  {is} determined by the crystallographic orientation of the lattice.
Here, we demonstrate theoretically within the Fr\"ohlich approach
and experimentally by photoluminescence emission that the bare trion  {binding energy} (20 meV) is significantly enhanced by the phonons at 
the interface between the two-dimensional semiconductor MoS$_2$
and the bulk transition metal oxide SrTiO$_3$.
The low-temperature  {binding energy} changes from 60 meV in [001]-oriented 
substrates to 90 meV for [111] orientation,
as a result of the counter-intuitive interplay between the rotational axis of the
MoS$_2$ trion and that of the SrTiO$_3$ phonon mode.


\end{abstract}


\maketitle

\begin{figure}
 \includegraphics[width=\columnwidth]{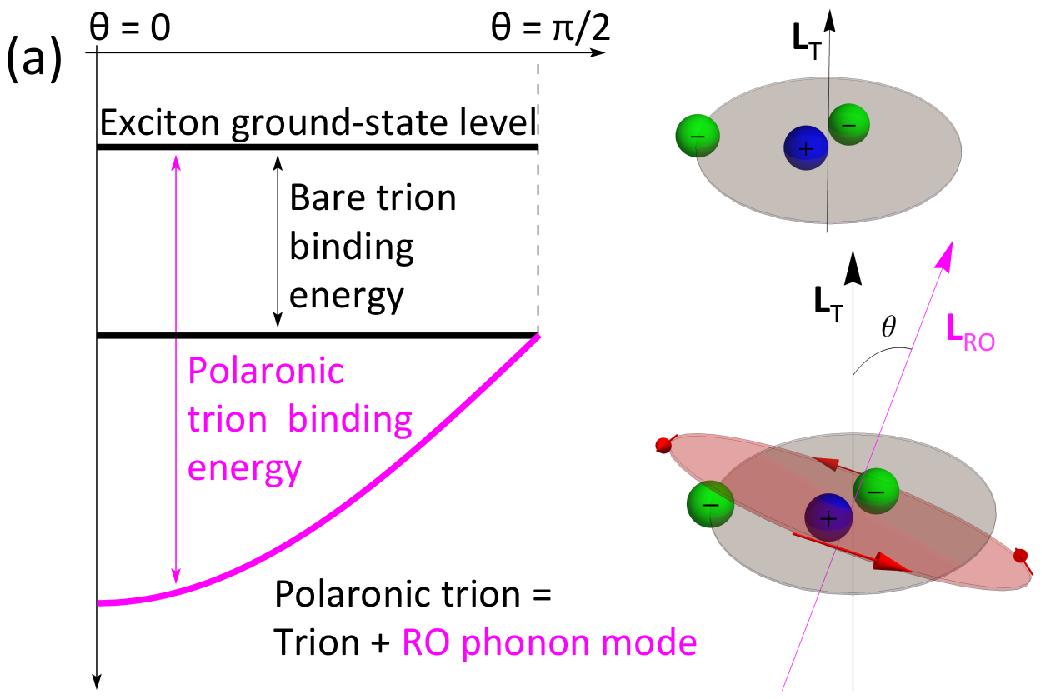}
 \includegraphics[width=\columnwidth]{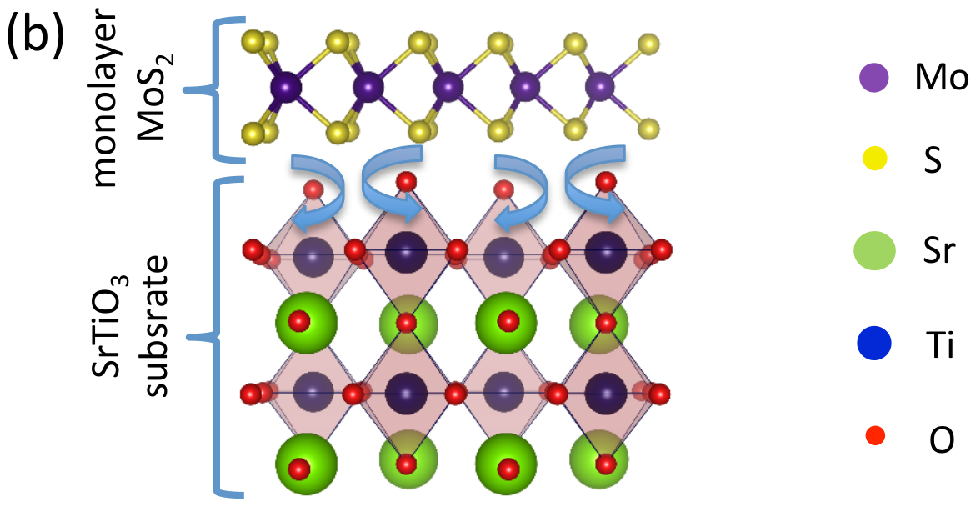}
 \caption{(a) 
A trion consists of a tightly-bound excitonic core 
and an electron weakly coupled to the core by the electron-dipole interactions.
In a polaronic trion, a RO phonon mode couples with the tangential momentum of the outer electron 
increasing the resulting quasiparticle BE.
The electrons and holes are represented by the green and blue balls, respectively.
The tangential polarization generated by the RO phonon mode is shown by the red arrows.
The black and magenta arrows show directions of the trion ($\mathbf{L}_\mathrm{T}$) and RO phonon ($\mathbf{L}_\mathrm{RO}$) angular momenta respectively,
and $\theta$ is the angle between them.
The trion-phonon coupling maximizes at $\theta=0$ and vanishes at $\theta=\pi/2$,
as shown by the polaronic trion BE curve (magenta).
The bare trion BE does not depend on $\theta$ at all.
(b) Schematic of MoS$_2$/ SrTiO$_3$ heterostructure utilized to create rotational Fr\"ohlich
interactions.  {MoS$_2$ being an $n$-type semiconductor exhibits negatively charged trions.
At low temperature, SrTiO$_3$ experiences structural cubic-to-tetragonal phase transition 
that activates the RO phonon mode due to the rotating TiO$_6$ octahedra (see also Fig. \ref{fig3}).}}
 \label{fig1}
\end{figure}

\begin{widetext}

\begin{figure}
\includegraphics[width=0.9\textwidth]{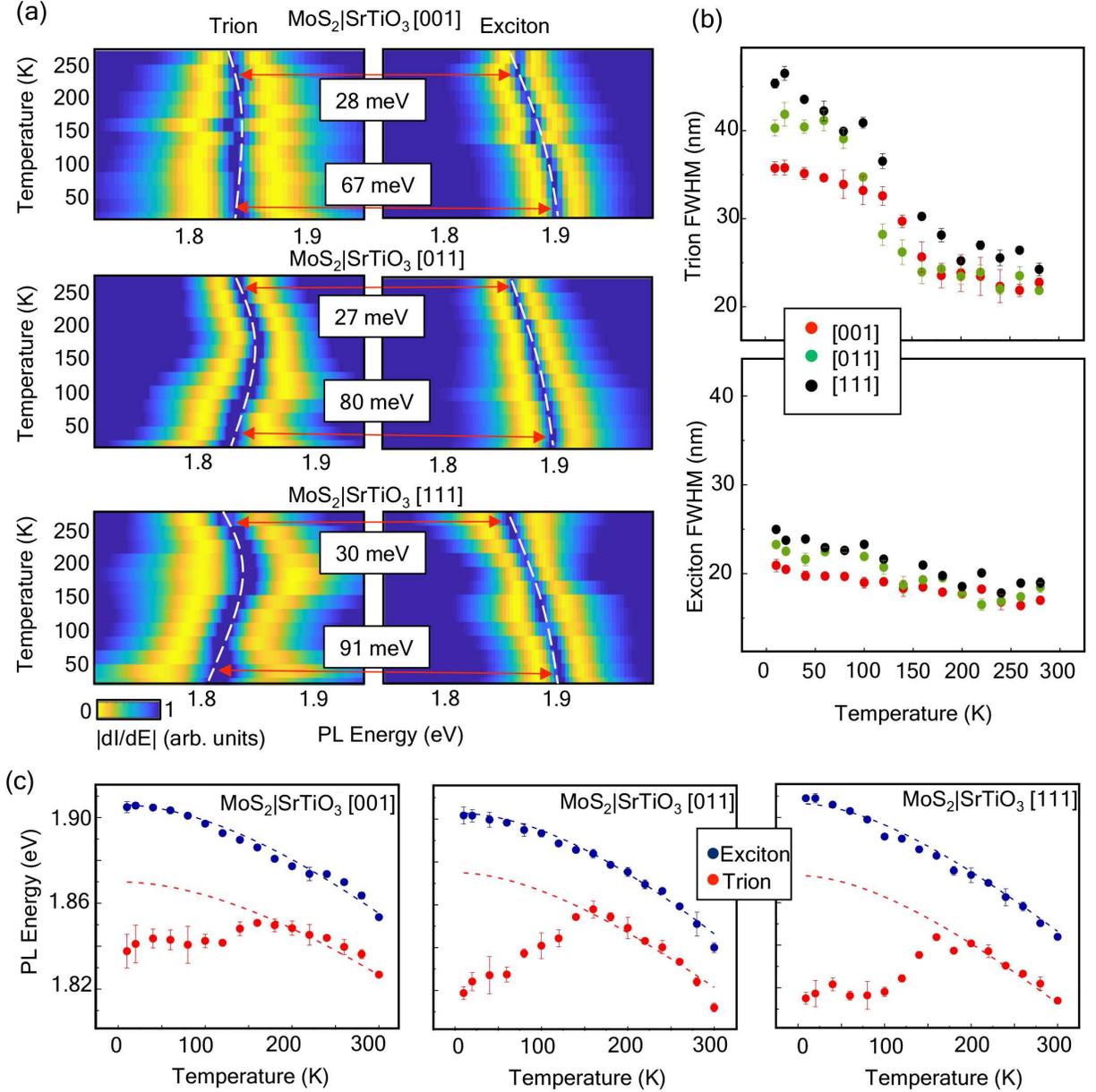}
\caption{(a) Pseudocolor map of the differential PL emission intensity 
($|dI/dE|$) from 2D MoS$_2$ demonstrates 
two quasiparticle peaks attributed to excitons and trions.
The splitting between them depends on temperature and SrTiO$_3$ substrate orientation
with the strongest separation for [111]-oriented SrTiO$_3$ crystals below 50 K.
(b) Temperature dependence of FWHM for trionic (upper panel) and excitonic (lower panel) PL quasiparticle peaks
for different SrTiO$_3$ substrate orientations indicates much stronger Fr\"ohlich interactions for the former than for the latter.
(c) Extracted trionic and excitonic PL energies vs. temperature with the corresponding Varshni fits as dashed lines
indicate anomalous behavior of the PL trion peak below 132 K.
The error bars are standard error for three samples.}
\label{fig2}
\end{figure}

\begin{figure}
\includegraphics[width=0.9\textwidth]{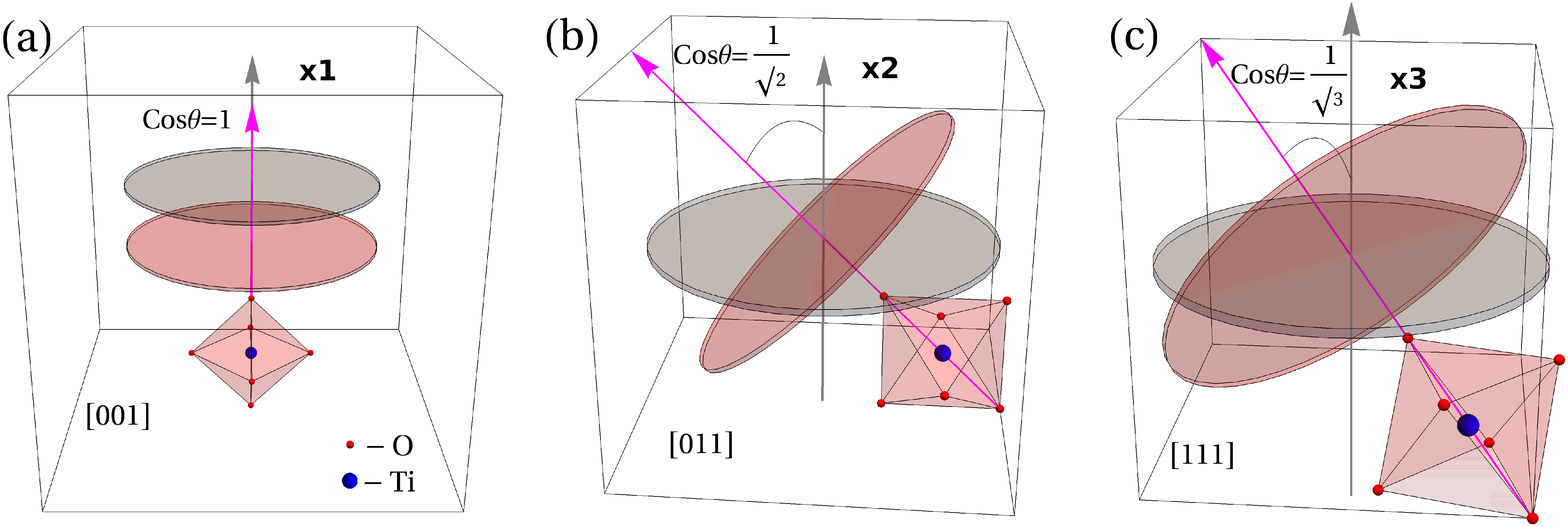}
\caption{(a) In a [001]-oriented SrTiO$_3$ domain, the TiO$_6$ octahedra rotate in the same plane as the trion in 2D MoS$_2$ placed on top.
Here, the trion's and RO phonon planes of rotation are shown by the gray
and reddish discs, respectively. 
(b,c) Once the SrTiO$_3$ substrate orientation changes, the rotational axes of the trion and RO phonon mode 
(depicted by black and magenta arrows, respectively) do not align anymore and span
the angle $\theta$ determined by the domain orientation.
The bold numbers indicate the number of mutually perpendicular domain orientations 
contributing to the polaronic trion BE.}
\label{fig3}
\end{figure}
 
\end{widetext}

{\em Introduction ---}
The quasiparticle concept is a powerful tool for understanding physics
of many-body phenomena \cite{quasiparticle-intro}.
The concept was invented a few decades ago to describe the Fermi liquid \cite{landau1957},
and later on applied to a broad range of phenomena including superconductivity \cite{bogoljubov1958,valatin1958}, 
magnetic ordering \cite{magnons-review-2015}, and fractional quantum Hall effect \cite{laughlin1999}.
The optically excited two-dimensional (2D) semiconductors contain tightly-bound excitons and trions
--- the quasiparticles composed of electrons and holes glued together by Coulomb forces \cite{photonics-review-2016}.
Recently, yet another exhibit in this quasiparticle zoo
--- a {\em polaronic trion} was reported \cite{AdvMat2019soumya}.
In this Letter, we demonstrate that a rotational optical (RO) phonon mode is necessary for the trion to engage in polaronic coupling that explains the underlying mechanism leading to formation of polaronic trions  {and enables significant tunability of trion binding energy (BE), a key to realize trion based optoelectronics.}

In quasiparticle language, the conventional (Fr\"ohlich \cite{froelich1954,PR1955feynman}) polaron is an electron dressed with phonons. 
The energy needed to undress the polaron (i.e. to release the electron) is the polaron BE.
Typically, the strongest Fr\"ohlich coupling occurs with longitudinal optical (LO) phonons
in polar crystals with large difference between 
the static dielectric permittivity and its electronic contribution,
such as in SrTiO$_3$ \cite{devreese1996polarons,STOfrequency2006}.
However, the trion-phonon interaction is distinct from coupling of phonons to free electrons.
The outer electron in the trion is bound to the excitonic core
(see Fig. \ref{fig1}),
resulting in a finite angular momentum which enables stronger coupling with 
RO rather than LO phonon modes.
To maximize the effect, the trion's plane of rotation must match the polarization
plane created by the RO mode (see Fig. \ref{fig1}).
Hence, we can probe polaronic trions by either changing the Fr\"ohlich coupling itself or
the angle between rotational planes of the trion and the RO phonon mode.
 SrTiO$_3$ hosts RO phonons with very low vibration frequency \cite{STO-soft-phonon-temprature2001} enabling an ideal environment to investigate the rotational Fr\"ohlich coupling with trions. Notably, by changing the SrTiO$_3$ crystal orientation, one can tilt the rotational axis of 
the RO phonon mode and hence investigate the angular dependence of this coupling. 

{\em PL spectroscopy. ---}
Monolayer MoS$_2$ is grown on single crystal SrTiO$_3$ substrates
by chemical vapor deposition \cite{NatMat2013growth}, and our samples 
are of comparable quality with those reported previously, see  Supplemental Material \cite{suppl-info2} for sample characterization.
We use three different crystallographic orientations of the SrTiO$_3$ substrate
to tailor the polaronic effects in 2D MoS$_2$.
Fig. \ref{fig2}a shows the differential PL emission spectra of the excitonic (right) and trionic (left) peaks in the MoS$_2$ PL extracted from Lorentz fitting described in \cite{suppl-info2}. We have confirmed that the low energy peak is indeed a trion and does not arise from defect bound excitons through excitation power and electrostatic doping dependent measurements (Figs. S4  and S5 in \cite{suppl-info2}). The exciton-trion peak separation is the
trion BE we are after.  At low temperatures, the trionic peak
splits further away from the excitonic peak position, 
and the splitting turns out to be dependent on
the crystallographic orientation of the SrTiO$_3$ substrate.
We have achieved BE enhancement of up 
to $60$ meV that is enormous having in mind that bare trion BE
is less than $30$ meV. 
Fig. \ref{fig2}b shows the full width at half maximum (FWHM) for trionic and excitonic
PL peaks. The trion FWHM experiences a significant broadening below the soft phonon activation temperature
 \cite{STO-soft-phonon-temprature2001,STO-soft-phonon-temprature1968} $T_a\sim 132$ K,
and the broadening turns out to be strongly dependent on the substrate orientation.
In contrast, the exciton FWHM demonstrates much smaller broadening and weaker dependence on the substrate orientation.
Finally, the exciton emission energy exhibits the usual monotonic blue shift given by
the Varshni relation \cite{VARSHNI1967149} whereas the trion emission energy
undergoes an unusual red shift below $T_a$ (see Fig. \ref{fig2}c and Table S1 in \cite{suppl-info2}).
The data presented in Fig. \ref{fig2} all together indicates that the trion is not
a conventional trion anymore but is an entirely new quasiparticle, the polaronic trion \cite{AdvMat2019soumya}.

{\em Trion Hamiltonian. ---}
The trion can be seen as an electron weakly interacting with 
the excitonic core.
The unperturbed Hamiltonian describing the relative electron-exciton motion can be written as
\begin{equation}
\hat{H}_0= 
-\frac{\hbar^2}{2\mu_T}\left[\frac{1}{r^2}\frac{\partial^2}{\partial \varphi^2}
 +\frac{1}{r}\frac{\partial}{\partial r}\left( r \frac{\partial}{\partial r}\right)\right].
 \label{trion}
\end{equation}
Here, we use polar coordinates $\{\varphi,r\}$.
The trion reduced mass, $\mu_T=m_X m_e/m_T$, is  defined in terms of 
the trion ($m_T=m_X+m_e$), exciton ($m_X=m_e+m_h$), electron ($m_e$), and hole ($m_h$) effective masses, respectively.
The energy in Eq.~(\ref{trion}) is counted from the exciton ground-state level,
as shown in Fig. \ref{fig1}a.
The first (second) term in the square brackets is the tangential (radial) momentum operator
with the eigenvalues $k_\varphi$ ($k_r$) given in units of the Planck constant $\hbar$.
The electron-dipole interaction perturbing $\hat{H}_0$
is much weaker than the direct Coulomb potential 
responsible for the exciton formation and rapidly vanishes at the distances 
much larger than the exciton size.
This results in the trion BE much lower than that of exciton.

{\em Rotational Fr\"ohlich coupling. ---} 
The 2D Fourier transform of the polaronic interaction can be written 
as $|V_q|^2=8\pi^2 e^2 F^2/q$, where $e$ is the elementary charge,
$\mathbf{q}$ is the in-plane wave vector, and
$F$ is a proportionality coefficient between the phonon mode amplitude
and dielectric polarization created by this mode \cite{kittel1987quantum}.
To express $F$ in terms of macroscopic quantities 
we adopt an argument by Kittel~\cite{kittel1987quantum} where the phonon perturbation producing
dielectric polarization is equivalent to the Coulomb potential screened by the 
phononic part of the dielectric permittivity, {\em i.e.}
\begin{equation}
\frac{2e^2}{\hbar\omega}\sum_\mathbf{Q} \frac{(4\pi F)^2}{Q^2} \mathrm{e}^{i\mathbf{Q}\cdot\mathbf{r}} =
\left(\frac{1}{\epsilon_\infty}-\frac{1}{\epsilon_0}\right) 
 \sum_\mathbf{q} \frac{2\pi e^2}{q} \mathrm{e}^{i\mathbf{q}\cdot\mathbf{r}},
 \label{kittel}
\end{equation}
where $\epsilon_0$ ($\epsilon_\infty$) 
is the static (high-frequency) dielectric permittivity at the MoS$_2$/SrTiO$_3$ interface,
$\hbar \omega$ is the phonon energy quantum,
$\mathbf{r}$ is the in-plane coordinate,
$\mathbf{Q}$ is the phonon wave vector whose absolute value can be written
in terms of in-plane ($q$) and axial ($q_\parallel$) components as
$Q=\sqrt{q_\parallel^2+q^2 - 2q q_\parallel\cos(\pi/2+\theta)}$.
The axial component does not contribute to rotational Fr\"ohlich coupling
and can be integrated out easily.
The resulting polarization turns out to be $\theta$ dependent,
$F=\sqrt{\hbar\omega\cos\theta\left(\epsilon_\infty^{-1}-\epsilon_0^{-1}\right)/(8\pi)}$,
and the 2D Fourier transform of the polaronic potential reads
\begin{equation}
 V_q = - i \hbar \omega \sqrt{\cos\theta} \sqrt{\frac{2\pi\alpha r_\omega}{q}},
 \label{Vq}
\end{equation}
where $r_\omega = \sqrt{\hbar/2\mu_T \omega}$ is the interaction length, and 
\begin{equation}
 \alpha = \frac{e^2}{2\hbar\omega r_\omega} \left(\frac{1}{\epsilon_\infty} - \frac{1}{\epsilon_0}\right)
\end{equation}
is the standard Fr\"ohlich coupling constant \cite{froelich1954}.  
The striking difference between the standard 2D polaronic interaction \cite{devreese1996polarons,kittel1987quantum} and Eq. (\ref{Vq})
is the $\sqrt{\cos\theta}$ pre-factor that occurs due to the special direction
singled out by the angular momentum of a RO phonon mode.
Note that the effective mass in $r_\omega$ is given by $\mu_T$ instead of $m_e$ as in the conventional case \cite{devreese1996polarons}.

{\em Polaronic perturbation. ---}
In the non-perturbed limit, when both dipole and polaronic perturbations vanish,
the plane-wave solution suggests the kinetic energy $E_\mathbf{k}$
of the relative electron-exciton motion be a sum 
the radial $\hbar^2 k_r^2 /2\mu_T$ and tangential $\hbar^2 k_\varphi^2 /2\mu_T$ terms.
The latter is quantized in any circularly symmetric potential,
however, we assume that the normalization length is long enough 
to justify integration instead of summation and map tangential and radial momenta
onto the Cartesian coordinates.  {We note that within the Fr\"ohlich approach, the angular momenta indicated in Fig. \ref{fig1} are quasiclassical quantities and not associated with the $s$ or $p$ quantum states.}
The perturbation theory suggests the following expression for the polaronic energy correction \cite{kittel1987quantum}
\begin{equation}
E_P =  - \int\frac{d^2 q }{(2\pi)^2}
\frac{|V_q|^2}{E_\mathbf{k} - E_{\mathbf{k}-\mathbf{q}}-\hbar\omega}.
\label{EP}
\end{equation}
We evaluate Eq.~(\ref{EP}) for the BE correction ($\mathbf{k}\to 0$). 
Despite the electron-exciton relative motion being 2D,
the rotational polaronic coupling is effectively 1D.  This is because 
RO phonon modes produce no radial polarization (hence, no radial electric field, see 
Fig. \ref{fig1}a), and, therefore, the energy difference $E_\mathbf{k} - E_{\mathbf{k}-\mathbf{q}}$
does not contain $q_r$. The BE correction can then be written as
\begin{eqnarray}
\nonumber E_P & =& \frac{2}{\pi} \int\limits_0^\infty dq_r\int\limits_0^\infty dq_\varphi
\frac{\hbar^2\omega^2 \alpha r_\omega \cos\theta}{\sqrt{q_\varphi^2 + q_r^2}} 
\frac{2\mu_T/\hbar^2}{q_\varphi^2+r_\omega^{-2}}\\
&=& \alpha\hbar\omega \cos\theta \ln\left(2r_\omega/a\right),
\label{EP2}
\end{eqnarray}
where $1/a$ is a momentum cut-off.  Similar to the conventional expression for the 2D polaron BE, $E_P=\frac{\pi}{2}\alpha\hbar\omega$, our result is linear in $\alpha$ (this also holds beyond perturbation theory, see e.g. Refs.~\cite{DasSarma1985,PRB1985exactpolaron2D,devreese1996polarons}) and linear in phonon energy $\hbar \omega$, setting
the scale of polaronic interactions.  However, Eq.~(\ref{EP2}) is different in two important ways: since the RO phonon modes are decoupled from both
the axial and radial electron motion, this results respectively in the $\cos\theta$ and $\ln\left(2r_\omega/a\right)$ pre-factors (the latter is a weak function of the order of unity and less important than the former).  The logarithmic divergence is a well-known property of the Fr\"ohlich coupling in a 1D limit \cite{peeters1D}.
The length $a$ is the lattice constant that determines the first Brillouin zone size in MoS$_2$.  If $q_r$ is retained in the denominator of Eq. (\ref{EP}), and $\theta$ is set to zero, we recover the conventional result.


{\em Discussion. ---}
The dressed trion BE is $E_{PT} = E_P + E_T$,
where $E_T$ is the bare trion binding energy. 
To make predictions regarding BE in realistic samples
the multi-domain structure of the SrTiO$_3$ substrate must be taken into account.
The axis of antiphase rotation of neighboring oxygen 
TiO$_6$ octahedra is different in each domain \cite{STO-review}.
We assume that the domain orientation is perfectly random,
so that any of three mutually perpendicular 
orientations are weighted equally in the BE calculation.
In the simplest case of the [001]-grown substrate, the rotational axis of RO mode
in [001]-oriented domain is normal to the trion plane and the polaronic effect is maximal
(see Fig. \ref{fig3}a).
The other two [010]- and [100]-oriented domains do not contribute at all
because the phonon mode rotation axis is parallel to the trion plane and $\theta=\pi/2$
in Eq. (\ref{EP2}).
Hence, the total $E_{PT}$ reads
\begin{equation}
 \label{001}
 E_{PT}[001] = E_T + \alpha \hbar\omega \ln\left(2r_\omega/a\right).
\end{equation}
In the case of either [011] or [101] domain orientation we have $\cos\theta = 1/\sqrt{2}$
(see Fig. \ref{fig3}b).
The [110]-oriented domains do not contribute here, and $E_{PT}$ reads
\begin{equation}
\label{011}
E_{PT}[011] = E_T + \sqrt{2}\alpha \hbar\omega \ln\left(2r_\omega/a\right).
\end{equation}
The [111] orientation suggests $\cos\theta = 1/\sqrt{3}$ (see Fig. \ref{fig3}c),
and all three possible mutually  perpendicular domain orientations do contribute equally. Hence, we have
\begin{equation}
\label{111}
E_{PT}[111] =  E_T + \sqrt{3}\alpha \hbar\omega\ln\left(2r_\omega/a\right).
\end{equation}

\begin{figure}
\includegraphics[width=\columnwidth]{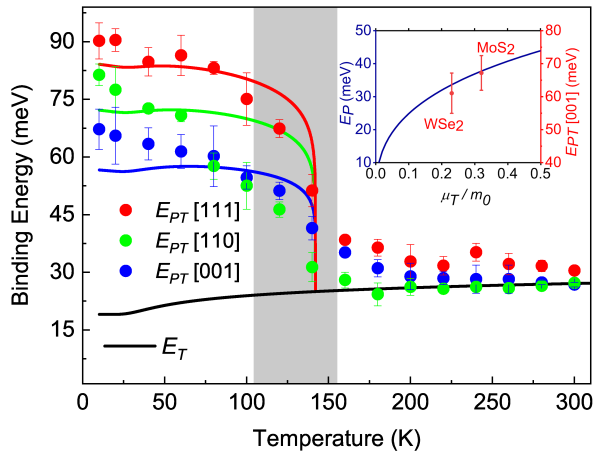}
\caption{Temperature dependence of the polaronic trion BE 
measured for three different substrate orientations along with  $E_{PT}$ given by Eqs. (\ref{001},\ref{011},\ref{111}), respectively.
The shaded transition region corresponds to $T_c^\mathrm{bulk}<T<T_c^\mathrm{surface}$
with $T_c^\mathrm{bulk}$ and $T_c^\mathrm{surface}$ being the bulk and surface structural transition temperatures, respectively \cite{STOdepth2011}. The model is not expected to fit BE in that region.
The bare trion energy~\cite{binding-energy-fit-2014} ($E_T$) is shown for comparison. The error bars are standard error for three samples.
Inset: Polaronic trion BE measured at 10 K for MoS$_2$ and WSe$_2$ 
on the same [001]-oriented SrTiO$_3$ substrate (right axis) follows
the theoretical trend predicted by Eq. (\ref{EP2}) at $T=0$ (left axis). We avoid plotting theoretical $E_{PT}$, as we do not know the phenomenological $E_T$ for WSe$_2$.}
\label{fig4}
\end{figure}

\noindent The BE is shown in Fig.~\ref{fig4} as a function of temperature
for different substrate orientations.  The highest BE ($\sim 90$ meV) is achieved for [111]-grown substrate despite the smallest $\cos\theta=1/\sqrt{3}$ (hence, the weakest coupling)
for [111]-domain orientation. This is almost twice larger
than the highest trion BE reported so far in n-doped MoS$_2$
($\sim 40$ meV \cite{binding-energy-fit-2014}, $\sim 50$ meV \cite{mak2013trions}).
Moreover, the polaronic trion BE allows for 
200\% tunability (from $\sim 30$ meV to $\sim 90$ meV),
which far exceeds that possible by conventional electrostatic gating \cite{mak2013trions}.  The bare trion energy $E_T$ (black line) is estimated using the phenomenological approach of Ref.~\cite{binding-energy-fit-2014}. 
Eqs. (\ref{001},\ref{011},\ref{111}) 
combined with the material parameters, $\epsilon_0(T)$, $\epsilon_\infty$, and $\omega(T)$,
are able to explain the measured polaronic trion BE behaviour
for all crystallographic orientations of the substrate.  We do not adjust the material parameters, see Supplemental Material \cite{suppl-info2}, which also includes Refs. \cite{NatMat2013growth,efimkin2017,binding-energy-fit-2014,STOdielectric1962,STO-dielectric-formula1952,2DMoS2dielectric2018,STO-high-freq-diel-1997,STO-soft-phonon-temprature2001,STO-soft-phonon-temprature1968,STOdepth2011,PRB2013berkelbach,lu2014van,eff-mass2016,lattice-constant,lee2010anomalous,mak2013trions,VARSHNI1967149,tongay2013defects,zhao2013lattice,WSe2-he,chellappan2018effect,jones2013optical}.

The polaronic effect we have discovered is not limited to trions in MoS$_2$. 
We have tested a few similar samples where MoS$_2$ has been substituted by
WSe$_2$ and found the same PL features as shown in Fig. \ref{fig2}. The PL emission from WSe$_2$/SrTiO$_3$ at 10 K (see Fig. S7 in \cite{suppl-info2}) shows spectral broadening and an anomalous enhancement in the energy separation between the exciton and trion peak (as compared with the spectra at 300 K), suggesting an increased trion binding energy.  
The polaronic trion BE in WSe$_2$ has turned out to be a little lower than in MoS$_2$,
see inset in Fig. \ref{fig4} for comparison.
As the high-frequency dielectric permittivity 
is the same for both MoS$_2$ and WSe$_2$ \cite{2DMoS2dielectric2018},
we attribute this difference to the lower effective mass in WSe$_2$ \cite{PRB2013berkelbach,eff-mass2016}, see \cite{suppl-info2} for details.

{\em Outlook. ---}
The polaronic trion discussed here is a complex three-component quasiparticle comprising the exciton core, an electron, and an RO phonon mode coupled together 
by both Coulomb and Fr\"ohlich interactions, resulting in a large enhancement
of the BE.  Since the polaronic interaction length $r_\omega$ is much larger than the lattice constant, we expect these polaronic trions to only weakly depend on SrTiO$_3$ surface termination \cite{schrott2001site}. Despite its complexity we have shown that the quasiparticle can live in 2D semiconductors other than MoS$_2$. We therefore anticipate further discoveries 
revealing a hierarchy of energy-rich quasiparticles \cite{barbone2018complexes} optically excited in 2D semiconductors with unconventional substrates underneath. 
 
{\em Acknowledgements. ---}
M.T. is supported by the Director's Senior Research Fellowship at CA2DM
(Singapore NRF Medium-Sized Centre Programme R-723-000-001-281,
NUS Young Investigator Award R-607-000-236-133). 
S.S., S.G., and T.V. acknowledge support from the National Research Foundation, Singapore under Competitive Research Program (NRF2015NRF-CRP001-015).
S.M. acknowledges support from NUSNNI general purpose account grant IN-398-000-006-001 and international atomic energy agency (IAEA) CRP (F11020). 

{\em Author contributions. ---}
M.T. and S.S contributed equally. M.T. devised the model. S.S. grew the MoS$_2$ samples on SrTiO$_3$ with the help of S.M. and performed the PL spectroscopy.  {P.S., Y.W., J.Y., W.L., J.L.M. and M.C. provided the WSe$_2$/SrTiO$_3$ samples. All the authors discussed and analysed the data. T.V. and S.A. supervised the project.}

$ ^*$ corresponding authors.

$ ^\dagger$ present address: Department of Physics, 
S. B. College, Mahatma Gandhi University, 
Kerala 686101 India.

\bibliography{trions.bib}

\begin{thebibliography}{42}%
\makeatletter
\providecommand \@ifxundefined [1]{%
 \@ifx{#1\undefined}
}%
\providecommand \@ifnum [1]{%
 \ifnum #1\expandafter \@firstoftwo
 \else \expandafter \@secondoftwo
 \fi
}%
\providecommand \@ifx [1]{%
 \ifx #1\expandafter \@firstoftwo
 \else \expandafter \@secondoftwo
 \fi
}%
\providecommand \natexlab [1]{#1}%
\providecommand \enquote  [1]{``#1''}%
\providecommand \bibnamefont  [1]{#1}%
\providecommand \bibfnamefont [1]{#1}%
\providecommand \citenamefont [1]{#1}%
\providecommand \href@noop [0]{\@secondoftwo}%
\providecommand \href [0]{\begingroup \@sanitize@url \@href}%
\providecommand \@href[1]{\@@startlink{#1}\@@href}%
\providecommand \@@href[1]{\endgroup#1\@@endlink}%
\providecommand \@sanitize@url [0]{\catcode `\\12\catcode `\$12\catcode
  `\&12\catcode `\#12\catcode `\^12\catcode `\_12\catcode `\%12\relax}%
\providecommand \@@startlink[1]{}%
\providecommand \@@endlink[0]{}%
\providecommand \url  [0]{\begingroup\@sanitize@url \@url }%
\providecommand \@url [1]{\endgroup\@href {#1}{\urlprefix }}%
\providecommand \urlprefix  [0]{URL }%
\providecommand \Eprint [0]{\href }%
\providecommand \doibase [0]{http://dx.doi.org/}%
\providecommand \selectlanguage [0]{\@gobble}%
\providecommand \bibinfo  [0]{\@secondoftwo}%
\providecommand \bibfield  [0]{\@secondoftwo}%
\providecommand \translation [1]{[#1]}%
\providecommand \BibitemOpen [0]{}%
\providecommand \bibitemStop [0]{}%
\providecommand \bibitemNoStop [0]{.\EOS\space}%
\providecommand \EOS [0]{\spacefactor3000\relax}%
\providecommand \BibitemShut  [1]{\csname bibitem#1\endcsname}%
\let\auto@bib@innerbib\@empty
\bibitem [{\citenamefont {Venema}\ \emph {et~al.}(2016)\citenamefont {Venema},
  \citenamefont {Verberck}, \citenamefont {Georgescu}, \citenamefont {Prando},
  \citenamefont {Couderc}, \citenamefont {Milana}, \citenamefont {Maragkou},
  \citenamefont {Persechini}, \citenamefont {Pacchioni},\ and\ \citenamefont
  {Fleet}}]{quasiparticle-intro}%
  \BibitemOpen
  \bibfield  {author} {\bibinfo {author} {\bibfnamefont {L.}~\bibnamefont
  {Venema}}, \bibinfo {author} {\bibfnamefont {B.}~\bibnamefont {Verberck}},
  \bibinfo {author} {\bibfnamefont {I.}~\bibnamefont {Georgescu}}, \bibinfo
  {author} {\bibfnamefont {G.}~\bibnamefont {Prando}}, \bibinfo {author}
  {\bibfnamefont {E.}~\bibnamefont {Couderc}}, \bibinfo {author} {\bibfnamefont
  {S.}~\bibnamefont {Milana}}, \bibinfo {author} {\bibfnamefont
  {M.}~\bibnamefont {Maragkou}}, \bibinfo {author} {\bibfnamefont
  {L.}~\bibnamefont {Persechini}}, \bibinfo {author} {\bibfnamefont
  {G.}~\bibnamefont {Pacchioni}}, \ and\ \bibinfo {author} {\bibfnamefont
  {L.}~\bibnamefont {Fleet}},\ }\href@noop {} {\bibfield  {journal} {\bibinfo
  {journal} {Nature Physics}\ }\textbf {\bibinfo {volume} {12}},\ \bibinfo
  {pages} {1085} (\bibinfo {year} {2016})}\BibitemShut {NoStop}%
\bibitem [{\citenamefont {Landau}(1956)}]{landau1957}%
  \BibitemOpen
  \bibfield  {author} {\bibinfo {author} {\bibfnamefont {L.~D.}\ \bibnamefont
  {Landau}},\ }\href@noop {} {\bibfield  {journal} {\bibinfo  {journal} {Sov.
  Phys. JETP}\ }\textbf {\bibinfo {volume} {3}},\ \bibinfo {pages} {920}
  (\bibinfo {year} {1956})}\BibitemShut {NoStop}%
\bibitem [{\citenamefont {Bogoljubov}(1958)}]{bogoljubov1958}%
  \BibitemOpen
  \bibfield  {author} {\bibinfo {author} {\bibfnamefont {N.~N.}\ \bibnamefont
  {Bogoljubov}},\ }\href {\doibase 10.1007/BF02745585} {\bibfield  {journal}
  {\bibinfo  {journal} {Il Nuovo Cimento (1955-1965)}\ }\textbf {\bibinfo
  {volume} {7}},\ \bibinfo {pages} {794} (\bibinfo {year} {1958})}\BibitemShut
  {NoStop}%
\bibitem [{\citenamefont {Valatin}(1958)}]{valatin1958}%
  \BibitemOpen
  \bibfield  {author} {\bibinfo {author} {\bibfnamefont {J.~G.}\ \bibnamefont
  {Valatin}},\ }\href {\doibase 10.1007/BF02745589} {\bibfield  {journal}
  {\bibinfo  {journal} {Il Nuovo Cimento (1955-1965)}\ }\textbf {\bibinfo
  {volume} {7}},\ \bibinfo {pages} {843} (\bibinfo {year} {1958})}\BibitemShut
  {NoStop}%
\bibitem [{\citenamefont {Chumak}\ \emph {et~al.}(2015)\citenamefont {Chumak},
  \citenamefont {Vasyuchka}, \citenamefont {Serga},\ and\ \citenamefont
  {Hillebrands}}]{magnons-review-2015}%
  \BibitemOpen
  \bibfield  {author} {\bibinfo {author} {\bibfnamefont {A.~V.}\ \bibnamefont
  {Chumak}}, \bibinfo {author} {\bibfnamefont {V.~I.}\ \bibnamefont
  {Vasyuchka}}, \bibinfo {author} {\bibfnamefont {A.~A.}\ \bibnamefont
  {Serga}}, \ and\ \bibinfo {author} {\bibfnamefont {B.}~\bibnamefont
  {Hillebrands}},\ }\href@noop {} {\bibfield  {journal} {\bibinfo  {journal}
  {Nature Physics}\ }\textbf {\bibinfo {volume} {11}},\ \bibinfo {pages} {453}
  (\bibinfo {year} {2015})}\BibitemShut {NoStop}%
\bibitem [{\citenamefont {Laughlin}(1999)}]{laughlin1999}%
  \BibitemOpen
  \bibfield  {author} {\bibinfo {author} {\bibfnamefont {R.~B.}\ \bibnamefont
  {Laughlin}},\ }\href {\doibase 10.1103/RevModPhys.71.863} {\bibfield
  {journal} {\bibinfo  {journal} {Rev. Mod. Phys.}\ }\textbf {\bibinfo {volume}
  {71}},\ \bibinfo {pages} {863} (\bibinfo {year} {1999})}\BibitemShut
  {NoStop}%
\bibitem [{\citenamefont {Mak}\ and\ \citenamefont
  {Shan}(2016)}]{photonics-review-2016}%
  \BibitemOpen
  \bibfield  {author} {\bibinfo {author} {\bibfnamefont {K.~F.}\ \bibnamefont
  {Mak}}\ and\ \bibinfo {author} {\bibfnamefont {J.}~\bibnamefont {Shan}},\
  }\href@noop {} {\bibfield  {journal} {\bibinfo  {journal} {Nature Photonics}\
  }\textbf {\bibinfo {volume} {10}},\ \bibinfo {pages} {216} (\bibinfo {year}
  {2016})}\BibitemShut {NoStop}%
\bibitem [{\citenamefont {Sarkar}\ \emph {et~al.}(2019)\citenamefont {Sarkar},
  \citenamefont {Goswami}, \citenamefont {Trushin}, \citenamefont {Saha},
  \citenamefont {Panahandeh-Fard}, \citenamefont {Prakash}, \citenamefont
  {Tan}, \citenamefont {Scott}, \citenamefont {Loh}, \citenamefont {Adam},
  \citenamefont {Mathew},\ and\ \citenamefont {Venkatesan}}]{AdvMat2019soumya}%
  \BibitemOpen
  \bibfield  {author} {\bibinfo {author} {\bibfnamefont {S.}~\bibnamefont
  {Sarkar}}, \bibinfo {author} {\bibfnamefont {S.}~\bibnamefont {Goswami}},
  \bibinfo {author} {\bibfnamefont {M.}~\bibnamefont {Trushin}}, \bibinfo
  {author} {\bibfnamefont {S.}~\bibnamefont {Saha}}, \bibinfo {author}
  {\bibfnamefont {M.}~\bibnamefont {Panahandeh-Fard}}, \bibinfo {author}
  {\bibfnamefont {S.}~\bibnamefont {Prakash}}, \bibinfo {author} {\bibfnamefont
  {S.~J.~R.}\ \bibnamefont {Tan}}, \bibinfo {author} {\bibfnamefont
  {M.}~\bibnamefont {Scott}}, \bibinfo {author} {\bibfnamefont {K.~P.}\
  \bibnamefont {Loh}}, \bibinfo {author} {\bibfnamefont {S.}~\bibnamefont
  {Adam}}, \bibinfo {author} {\bibfnamefont {S.}~\bibnamefont {Mathew}}, \ and\
  \bibinfo {author} {\bibfnamefont {T.}~\bibnamefont {Venkatesan}},\ }\href
  {\doibase 10.1002/adma.201903569} {\bibfield  {journal} {\bibinfo  {journal}
  {Advanced Materials}\ }\textbf {\bibinfo {volume} {31}},\ \bibinfo {pages}
  {1903569} (\bibinfo {year} {2019})}\BibitemShut {NoStop}%
\bibitem [{\citenamefont {Fr\"ohlich}(1954)}]{froelich1954}%
  \BibitemOpen
  \bibfield  {author} {\bibinfo {author} {\bibfnamefont {H.}~\bibnamefont
  {Fr\"ohlich}},\ }\href {\doibase 10.1080/00018735400101213} {\bibfield
  {journal} {\bibinfo  {journal} {Advances in Physics}\ }\textbf {\bibinfo
  {volume} {3}},\ \bibinfo {pages} {325} (\bibinfo {year} {1954})}\BibitemShut
  {NoStop}%
\bibitem [{\citenamefont {Feynman}(1955)}]{PR1955feynman}%
  \BibitemOpen
  \bibfield  {author} {\bibinfo {author} {\bibfnamefont {R.~P.}\ \bibnamefont
  {Feynman}},\ }\href {\doibase 10.1103/PhysRev.97.660} {\bibfield  {journal}
  {\bibinfo  {journal} {Phys. Rev.}\ }\textbf {\bibinfo {volume} {97}},\
  \bibinfo {pages} {660} (\bibinfo {year} {1955})}\BibitemShut {NoStop}%
\bibitem [{\citenamefont {Devreese}(1996)}]{devreese1996polarons}%
  \BibitemOpen
  \bibfield  {author} {\bibinfo {author} {\bibfnamefont {J.~T.}\ \bibnamefont
  {Devreese}},\ }\href@noop {} {\bibfield  {journal} {\bibinfo  {journal}
  {Encycl. Appl. Phys.}\ }\textbf {\bibinfo {volume} {14}},\ \bibinfo {pages}
  {383} (\bibinfo {year} {1996})}\BibitemShut {NoStop}%
\bibitem [{\citenamefont {Takashima}\ \emph {et~al.}(2006)\citenamefont
  {Takashima}, \citenamefont {Wang}, \citenamefont {Prijamboedi}, \citenamefont
  {Shoji},\ and\ \citenamefont {Itoh}}]{STOfrequency2006}%
  \BibitemOpen
  \bibfield  {author} {\bibinfo {author} {\bibfnamefont {H.}~\bibnamefont
  {Takashima}}, \bibinfo {author} {\bibfnamefont {R.}~\bibnamefont {Wang}},
  \bibinfo {author} {\bibfnamefont {B.}~\bibnamefont {Prijamboedi}}, \bibinfo
  {author} {\bibfnamefont {A.}~\bibnamefont {Shoji}}, \ and\ \bibinfo {author}
  {\bibfnamefont {M.}~\bibnamefont {Itoh}},\ }\href {\doibase
  10.1080/00150190600689217} {\bibfield  {journal} {\bibinfo  {journal}
  {Ferroelectrics}\ }\textbf {\bibinfo {volume} {335}},\ \bibinfo {pages} {45}
  (\bibinfo {year} {2006})}\BibitemShut {NoStop}%
\bibitem [{\citenamefont {Petzelt}\ \emph {et~al.}(2001)\citenamefont
  {Petzelt}, \citenamefont {Ostapchuk}, \citenamefont {Gregora}, \citenamefont
  {Rychetsk\'y}, \citenamefont {Hoffmann-Eifert}, \citenamefont {Pronin},
  \citenamefont {Yuzyuk}, \citenamefont {Gorshunov}, \citenamefont {Kamba},
  \citenamefont {Bovtun}, \citenamefont {Pokorn\'y}, \citenamefont {Savinov},
  \citenamefont {Porokhonskyy}, \citenamefont {Rafaja}, \citenamefont
  {Van\ifmmode~\check{e}\else \v{e}\fi{}k}, \citenamefont {Almeida},
  \citenamefont {Chaves}, \citenamefont {Volkov}, \citenamefont {Dressel},\
  and\ \citenamefont {Waser}}]{STO-soft-phonon-temprature2001}%
  \BibitemOpen
  \bibfield  {author} {\bibinfo {author} {\bibfnamefont {J.}~\bibnamefont
  {Petzelt}}, \bibinfo {author} {\bibfnamefont {T.}~\bibnamefont {Ostapchuk}},
  \bibinfo {author} {\bibfnamefont {I.}~\bibnamefont {Gregora}}, \bibinfo
  {author} {\bibfnamefont {I.}~\bibnamefont {Rychetsk\'y}}, \bibinfo {author}
  {\bibfnamefont {S.}~\bibnamefont {Hoffmann-Eifert}}, \bibinfo {author}
  {\bibfnamefont {A.~V.}\ \bibnamefont {Pronin}}, \bibinfo {author}
  {\bibfnamefont {Y.}~\bibnamefont {Yuzyuk}}, \bibinfo {author} {\bibfnamefont
  {B.~P.}\ \bibnamefont {Gorshunov}}, \bibinfo {author} {\bibfnamefont
  {S.}~\bibnamefont {Kamba}}, \bibinfo {author} {\bibfnamefont
  {V.}~\bibnamefont {Bovtun}}, \bibinfo {author} {\bibfnamefont
  {J.}~\bibnamefont {Pokorn\'y}}, \bibinfo {author} {\bibfnamefont
  {M.}~\bibnamefont {Savinov}}, \bibinfo {author} {\bibfnamefont
  {V.}~\bibnamefont {Porokhonskyy}}, \bibinfo {author} {\bibfnamefont
  {D.}~\bibnamefont {Rafaja}}, \bibinfo {author} {\bibfnamefont
  {P.}~\bibnamefont {Van\ifmmode~\check{e}\else \v{e}\fi{}k}}, \bibinfo
  {author} {\bibfnamefont {A.}~\bibnamefont {Almeida}}, \bibinfo {author}
  {\bibfnamefont {M.~R.}\ \bibnamefont {Chaves}}, \bibinfo {author}
  {\bibfnamefont {A.~A.}\ \bibnamefont {Volkov}}, \bibinfo {author}
  {\bibfnamefont {M.}~\bibnamefont {Dressel}}, \ and\ \bibinfo {author}
  {\bibfnamefont {R.}~\bibnamefont {Waser}},\ }\href {\doibase
  10.1103/PhysRevB.64.184111} {\bibfield  {journal} {\bibinfo  {journal} {Phys.
  Rev. B}\ }\textbf {\bibinfo {volume} {64}},\ \bibinfo {pages} {184111}
  (\bibinfo {year} {2001})}\BibitemShut {NoStop}%
\bibitem [{\citenamefont {Van Der~Zande}\ \emph {et~al.}(2013)\citenamefont
  {Van Der~Zande}, \citenamefont {Huang}, \citenamefont {Chenet}, \citenamefont
  {Berkelbach}, \citenamefont {You}, \citenamefont {Lee}, \citenamefont
  {Heinz}, \citenamefont {Reichman}, \citenamefont {Muller},\ and\
  \citenamefont {Hone}}]{NatMat2013growth}%
  \BibitemOpen
  \bibfield  {author} {\bibinfo {author} {\bibfnamefont {A.~M.}\ \bibnamefont
  {Van Der~Zande}}, \bibinfo {author} {\bibfnamefont {P.~Y.}\ \bibnamefont
  {Huang}}, \bibinfo {author} {\bibfnamefont {D.~A.}\ \bibnamefont {Chenet}},
  \bibinfo {author} {\bibfnamefont {T.~C.}\ \bibnamefont {Berkelbach}},
  \bibinfo {author} {\bibfnamefont {Y.}~\bibnamefont {You}}, \bibinfo {author}
  {\bibfnamefont {G.-H.}\ \bibnamefont {Lee}}, \bibinfo {author} {\bibfnamefont
  {T.~F.}\ \bibnamefont {Heinz}}, \bibinfo {author} {\bibfnamefont {D.~R.}\
  \bibnamefont {Reichman}}, \bibinfo {author} {\bibfnamefont {D.~A.}\
  \bibnamefont {Muller}}, \ and\ \bibinfo {author} {\bibfnamefont {J.~C.}\
  \bibnamefont {Hone}},\ }\href@noop {} {\bibfield  {journal} {\bibinfo
  {journal} {Nature Materials}\ }\textbf {\bibinfo {volume} {12}},\ \bibinfo
  {pages} {554} (\bibinfo {year} {2013})}\BibitemShut {NoStop}%
\bibitem [{sup()}]{suppl-info2}%
  \BibitemOpen
  \href@noop {} {}\bibinfo {note} {Supplemental Material can be found
  below.}\BibitemShut {Stop}%
\bibitem [{\citenamefont {Fleury}\ \emph {et~al.}(1968)\citenamefont {Fleury},
  \citenamefont {Scott},\ and\ \citenamefont
  {Worlock}}]{STO-soft-phonon-temprature1968}%
  \BibitemOpen
  \bibfield  {author} {\bibinfo {author} {\bibfnamefont {P.~A.}\ \bibnamefont
  {Fleury}}, \bibinfo {author} {\bibfnamefont {J.~F.}\ \bibnamefont {Scott}}, \
  and\ \bibinfo {author} {\bibfnamefont {J.~M.}\ \bibnamefont {Worlock}},\
  }\href {\doibase 10.1103/PhysRevLett.21.16} {\bibfield  {journal} {\bibinfo
  {journal} {Phys. Rev. Lett.}\ }\textbf {\bibinfo {volume} {21}},\ \bibinfo
  {pages} {16} (\bibinfo {year} {1968})}\BibitemShut {NoStop}%
\bibitem [{\citenamefont {Varshni}(1967)}]{VARSHNI1967149}%
  \BibitemOpen
  \bibfield  {author} {\bibinfo {author} {\bibfnamefont {Y.}~\bibnamefont
  {Varshni}},\ }\href {\doibase https://doi.org/10.1016/0031-8914(67)90062-6}
  {\bibfield  {journal} {\bibinfo  {journal} {Physica}\ }\textbf {\bibinfo
  {volume} {34}},\ \bibinfo {pages} {149 } (\bibinfo {year}
  {1967})}\BibitemShut {NoStop}%
\bibitem [{\citenamefont {Kittel}(1987)}]{kittel1987quantum}%
  \BibitemOpen
  \bibfield  {author} {\bibinfo {author} {\bibfnamefont {C.}~\bibnamefont
  {Kittel}},\ }\href@noop {} {\emph {\bibinfo {title} {Quantum theory of
  solids}}}\ (\bibinfo  {publisher} {Wiley, New York},\ \bibinfo {year}
  {1987})\BibitemShut {NoStop}%
\bibitem [{\citenamefont {{Das~Sarma}}\ and\ \citenamefont
  {Mason}(1985)}]{DasSarma1985}%
  \BibitemOpen
  \bibfield  {author} {\bibinfo {author} {\bibfnamefont {S.}~\bibnamefont
  {{Das~Sarma}}}\ and\ \bibinfo {author} {\bibfnamefont {B.~A.}\ \bibnamefont
  {Mason}},\ }\href {\doibase https://doi.org/10.1016/0003-4916(85)90351-3}
  {\bibfield  {journal} {\bibinfo  {journal} {Annals of Physics}\ }\textbf
  {\bibinfo {volume} {163}},\ \bibinfo {pages} {78 } (\bibinfo {year}
  {1985})}\BibitemShut {NoStop}%
\bibitem [{\citenamefont {Xiaoguang}\ \emph {et~al.}(1985)\citenamefont
  {Xiaoguang}, \citenamefont {Peeters},\ and\ \citenamefont
  {Devreese}}]{PRB1985exactpolaron2D}%
  \BibitemOpen
  \bibfield  {author} {\bibinfo {author} {\bibfnamefont {W.}~\bibnamefont
  {Xiaoguang}}, \bibinfo {author} {\bibfnamefont {F.~M.}\ \bibnamefont
  {Peeters}}, \ and\ \bibinfo {author} {\bibfnamefont {J.~T.}\ \bibnamefont
  {Devreese}},\ }\href {\doibase 10.1103/PhysRevB.31.3420} {\bibfield
  {journal} {\bibinfo  {journal} {Phys. Rev. B}\ }\textbf {\bibinfo {volume}
  {31}},\ \bibinfo {pages} {3420} (\bibinfo {year} {1985})}\BibitemShut
  {NoStop}%
\bibitem [{\citenamefont {Peeters}\ and\ \citenamefont
  {Smondyrev}(1991)}]{peeters1D}%
  \BibitemOpen
  \bibfield  {author} {\bibinfo {author} {\bibfnamefont {F.~M.}\ \bibnamefont
  {Peeters}}\ and\ \bibinfo {author} {\bibfnamefont {M.~A.}\ \bibnamefont
  {Smondyrev}},\ }\href {\doibase 10.1103/PhysRevB.43.4920} {\bibfield
  {journal} {\bibinfo  {journal} {Phys. Rev. B}\ }\textbf {\bibinfo {volume}
  {43}},\ \bibinfo {pages} {4920} (\bibinfo {year} {1991})}\BibitemShut
  {NoStop}%
\bibitem [{\citenamefont {Pai}\ \emph {et~al.}(2018)\citenamefont {Pai},
  \citenamefont {Tylan-Tyler}, \citenamefont {Irvin},\ and\ \citenamefont
  {Levy}}]{STO-review}%
  \BibitemOpen
  \bibfield  {author} {\bibinfo {author} {\bibfnamefont {Y.-Y.}\ \bibnamefont
  {Pai}}, \bibinfo {author} {\bibfnamefont {A.}~\bibnamefont {Tylan-Tyler}},
  \bibinfo {author} {\bibfnamefont {P.}~\bibnamefont {Irvin}}, \ and\ \bibinfo
  {author} {\bibfnamefont {J.}~\bibnamefont {Levy}},\ }\href {\doibase
  10.1088/1361-6633/aa892d} {\bibfield  {journal} {\bibinfo  {journal} {Reports
  on Progress in Physics}\ }\textbf {\bibinfo {volume} {81}},\ \bibinfo {pages}
  {036503} (\bibinfo {year} {2018})}\BibitemShut {NoStop}%
\bibitem [{\citenamefont {Salman}\ \emph {et~al.}(2011)\citenamefont {Salman},
  \citenamefont {Smadella}, \citenamefont {MacFarlane}, \citenamefont
  {Patterson}, \citenamefont {Willmott}, \citenamefont {Chow}, \citenamefont
  {Hossain}, \citenamefont {Saadaoui}, \citenamefont {Wang},\ and\
  \citenamefont {Kiefl}}]{STOdepth2011}%
  \BibitemOpen
  \bibfield  {author} {\bibinfo {author} {\bibfnamefont {Z.}~\bibnamefont
  {Salman}}, \bibinfo {author} {\bibfnamefont {M.}~\bibnamefont {Smadella}},
  \bibinfo {author} {\bibfnamefont {W.~A.}\ \bibnamefont {MacFarlane}},
  \bibinfo {author} {\bibfnamefont {B.~D.}\ \bibnamefont {Patterson}}, \bibinfo
  {author} {\bibfnamefont {P.~R.}\ \bibnamefont {Willmott}}, \bibinfo {author}
  {\bibfnamefont {K.~H.}\ \bibnamefont {Chow}}, \bibinfo {author}
  {\bibfnamefont {M.~D.}\ \bibnamefont {Hossain}}, \bibinfo {author}
  {\bibfnamefont {H.}~\bibnamefont {Saadaoui}}, \bibinfo {author}
  {\bibfnamefont {D.}~\bibnamefont {Wang}}, \ and\ \bibinfo {author}
  {\bibfnamefont {R.~F.}\ \bibnamefont {Kiefl}},\ }\href {\doibase
  10.1103/PhysRevB.83.224112} {\bibfield  {journal} {\bibinfo  {journal} {Phys.
  Rev. B}\ }\textbf {\bibinfo {volume} {83}},\ \bibinfo {pages} {224112}
  (\bibinfo {year} {2011})}\BibitemShut {NoStop}%
\bibitem [{\citenamefont {Lin}\ \emph {et~al.}(2014)\citenamefont {Lin},
  \citenamefont {Ling}, \citenamefont {Yu}, \citenamefont {Huang},
  \citenamefont {Hsu}, \citenamefont {Lee}, \citenamefont {Kong}, \citenamefont
  {Dresselhaus},\ and\ \citenamefont {Palacios}}]{binding-energy-fit-2014}%
  \BibitemOpen
  \bibfield  {author} {\bibinfo {author} {\bibfnamefont {Y.}~\bibnamefont
  {Lin}}, \bibinfo {author} {\bibfnamefont {X.}~\bibnamefont {Ling}}, \bibinfo
  {author} {\bibfnamefont {L.}~\bibnamefont {Yu}}, \bibinfo {author}
  {\bibfnamefont {S.}~\bibnamefont {Huang}}, \bibinfo {author} {\bibfnamefont
  {A.~L.}\ \bibnamefont {Hsu}}, \bibinfo {author} {\bibfnamefont {Y.-H.}\
  \bibnamefont {Lee}}, \bibinfo {author} {\bibfnamefont {J.}~\bibnamefont
  {Kong}}, \bibinfo {author} {\bibfnamefont {M.~S.}\ \bibnamefont
  {Dresselhaus}}, \ and\ \bibinfo {author} {\bibfnamefont {T.}~\bibnamefont
  {Palacios}},\ }\href {\doibase 10.1021/nl501988y} {\bibfield  {journal}
  {\bibinfo  {journal} {Nano Letters}\ }\textbf {\bibinfo {volume} {14}},\
  \bibinfo {pages} {5569} (\bibinfo {year} {2014})}\BibitemShut {NoStop}%
\bibitem [{\citenamefont {Mak}\ \emph {et~al.}(2013)\citenamefont {Mak},
  \citenamefont {He}, \citenamefont {Lee}, \citenamefont {Lee}, \citenamefont
  {Hone}, \citenamefont {Heinz},\ and\ \citenamefont {Shan}}]{mak2013trions}%
  \BibitemOpen
  \bibfield  {author} {\bibinfo {author} {\bibfnamefont {K.~F.}\ \bibnamefont
  {Mak}}, \bibinfo {author} {\bibfnamefont {K.}~\bibnamefont {He}}, \bibinfo
  {author} {\bibfnamefont {C.}~\bibnamefont {Lee}}, \bibinfo {author}
  {\bibfnamefont {G.~H.}\ \bibnamefont {Lee}}, \bibinfo {author} {\bibfnamefont
  {J.}~\bibnamefont {Hone}}, \bibinfo {author} {\bibfnamefont {T.~F.}\
  \bibnamefont {Heinz}}, \ and\ \bibinfo {author} {\bibfnamefont
  {J.}~\bibnamefont {Shan}},\ }\href@noop {} {\bibfield  {journal} {\bibinfo
  {journal} {Nature Materials}\ }\textbf {\bibinfo {volume} {12}},\ \bibinfo
  {pages} {207} (\bibinfo {year} {2013})}\BibitemShut {NoStop}%
\bibitem [{\citenamefont {Efimkin}\ and\ \citenamefont
  {MacDonald}(2017)}]{efimkin2017}%
  \BibitemOpen
  \bibfield  {author} {\bibinfo {author} {\bibfnamefont {D.~K.}\ \bibnamefont
  {Efimkin}}\ and\ \bibinfo {author} {\bibfnamefont {A.~H.}\ \bibnamefont
  {MacDonald}},\ }\href {\doibase 10.1103/PhysRevB.95.035417} {\bibfield
  {journal} {\bibinfo  {journal} {Phys. Rev. B}\ }\textbf {\bibinfo {volume}
  {95}},\ \bibinfo {pages} {035417} (\bibinfo {year} {2017})}\BibitemShut
  {NoStop}%
\bibitem [{\citenamefont {Sawaguchi}\ \emph {et~al.}(1962)\citenamefont
  {Sawaguchi}, \citenamefont {Kikuchi},\ and\ \citenamefont
  {Kodera}}]{STOdielectric1962}%
  \BibitemOpen
  \bibfield  {author} {\bibinfo {author} {\bibfnamefont {E.}~\bibnamefont
  {Sawaguchi}}, \bibinfo {author} {\bibfnamefont {A.}~\bibnamefont {Kikuchi}},
  \ and\ \bibinfo {author} {\bibfnamefont {Y.}~\bibnamefont {Kodera}},\ }\href
  {\doibase 10.1143/JPSJ.17.1666} {\bibfield  {journal} {\bibinfo  {journal}
  {Journal of the Physical Society of Japan}\ }\textbf {\bibinfo {volume}
  {17}},\ \bibinfo {pages} {1666} (\bibinfo {year} {1962})}\BibitemShut
  {NoStop}%
\bibitem [{\citenamefont {Barrett}(1952)}]{STO-dielectric-formula1952}%
  \BibitemOpen
  \bibfield  {author} {\bibinfo {author} {\bibfnamefont {J.~H.}\ \bibnamefont
  {Barrett}},\ }\href {\doibase 10.1103/PhysRev.86.118} {\bibfield  {journal}
  {\bibinfo  {journal} {Phys. Rev.}\ }\textbf {\bibinfo {volume} {86}},\
  \bibinfo {pages} {118} (\bibinfo {year} {1952})}\BibitemShut {NoStop}%
\bibitem [{\citenamefont {Laturia}\ \emph {et~al.}(2018)\citenamefont
  {Laturia}, \citenamefont {Van~de Put},\ and\ \citenamefont
  {Vandenberghe}}]{2DMoS2dielectric2018}%
  \BibitemOpen
  \bibfield  {author} {\bibinfo {author} {\bibfnamefont {A.}~\bibnamefont
  {Laturia}}, \bibinfo {author} {\bibfnamefont {M.~L.}\ \bibnamefont {Van~de
  Put}}, \ and\ \bibinfo {author} {\bibfnamefont {W.~G.}\ \bibnamefont
  {Vandenberghe}},\ }\href@noop {} {\bibfield  {journal} {\bibinfo  {journal}
  {npj 2D Materials and Applications}\ }\textbf {\bibinfo {volume} {2}},\
  \bibinfo {pages} {6} (\bibinfo {year} {2018})}\BibitemShut {NoStop}%
\bibitem [{\citenamefont {Lasota}\ \emph {et~al.}(1997)\citenamefont {Lasota},
  \citenamefont {Wang}, \citenamefont {Yu},\ and\ \citenamefont
  {Krakauer}}]{STO-high-freq-diel-1997}%
  \BibitemOpen
  \bibfield  {author} {\bibinfo {author} {\bibfnamefont {C.}~\bibnamefont
  {Lasota}}, \bibinfo {author} {\bibfnamefont {C.-Z.}\ \bibnamefont {Wang}},
  \bibinfo {author} {\bibfnamefont {R.}~\bibnamefont {Yu}}, \ and\ \bibinfo
  {author} {\bibfnamefont {H.}~\bibnamefont {Krakauer}},\ }\href {\doibase
  10.1080/00150199708016086} {\bibfield  {journal} {\bibinfo  {journal}
  {Ferroelectrics}\ }\textbf {\bibinfo {volume} {194}},\ \bibinfo {pages} {109}
  (\bibinfo {year} {1997})}\BibitemShut {NoStop}%
\bibitem [{\citenamefont {Berkelbach}\ \emph {et~al.}(2013)\citenamefont
  {Berkelbach}, \citenamefont {Hybertsen},\ and\ \citenamefont
  {Reichman}}]{PRB2013berkelbach}%
  \BibitemOpen
  \bibfield  {author} {\bibinfo {author} {\bibfnamefont {T.~C.}\ \bibnamefont
  {Berkelbach}}, \bibinfo {author} {\bibfnamefont {M.~S.}\ \bibnamefont
  {Hybertsen}}, \ and\ \bibinfo {author} {\bibfnamefont {D.~R.}\ \bibnamefont
  {Reichman}},\ }\href {\doibase 10.1103/PhysRevB.88.045318} {\bibfield
  {journal} {\bibinfo  {journal} {Phys. Rev. B}\ }\textbf {\bibinfo {volume}
  {88}},\ \bibinfo {pages} {045318} (\bibinfo {year} {2013})}\BibitemShut
  {NoStop}%
\bibitem [{\citenamefont {Lu}\ \emph {et~al.}(2014)\citenamefont {Lu},
  \citenamefont {Guo}, \citenamefont {Wang}, \citenamefont {Wu},\ and\
  \citenamefont {Zeng}}]{lu2014van}%
  \BibitemOpen
  \bibfield  {author} {\bibinfo {author} {\bibfnamefont {N.}~\bibnamefont
  {Lu}}, \bibinfo {author} {\bibfnamefont {H.}~\bibnamefont {Guo}}, \bibinfo
  {author} {\bibfnamefont {L.}~\bibnamefont {Wang}}, \bibinfo {author}
  {\bibfnamefont {X.}~\bibnamefont {Wu}}, \ and\ \bibinfo {author}
  {\bibfnamefont {X.~C.}\ \bibnamefont {Zeng}},\ }\href@noop {} {\bibfield
  {journal} {\bibinfo  {journal} {Nanoscale}\ }\textbf {\bibinfo {volume}
  {6}},\ \bibinfo {pages} {4566} (\bibinfo {year} {2014})}\BibitemShut
  {NoStop}%
\bibitem [{\citenamefont {{Luisier}}\ \emph {et~al.}(2016)\citenamefont
  {{Luisier}}, \citenamefont {{Szabo}}, \citenamefont {{Stieger}},
  \citenamefont {{Klinkert}}, \citenamefont {{Br\"uck}}, \citenamefont
  {{Jain}},\ and\ \citenamefont {{Novotny}}}]{eff-mass2016}%
  \BibitemOpen
  \bibfield  {author} {\bibinfo {author} {\bibfnamefont {M.}~\bibnamefont
  {{Luisier}}}, \bibinfo {author} {\bibfnamefont {A.}~\bibnamefont {{Szabo}}},
  \bibinfo {author} {\bibfnamefont {C.}~\bibnamefont {{Stieger}}}, \bibinfo
  {author} {\bibfnamefont {C.}~\bibnamefont {{Klinkert}}}, \bibinfo {author}
  {\bibfnamefont {S.}~\bibnamefont {{Br\"uck}}}, \bibinfo {author}
  {\bibfnamefont {A.}~\bibnamefont {{Jain}}}, \ and\ \bibinfo {author}
  {\bibfnamefont {L.}~\bibnamefont {{Novotny}}},\ }in\ \href@noop {} {\emph
  {\bibinfo {booktitle} {2016 IEEE International Electron Devices Meeting
  (IEDM)}}}\ (\bibinfo {year} {2016})\ pp.\ \bibinfo {pages}
  {5.4.1--5.4.4}\BibitemShut {NoStop}%
\bibitem [{\citenamefont {Wakabayashi}\ \emph {et~al.}(1975)\citenamefont
  {Wakabayashi}, \citenamefont {Smith},\ and\ \citenamefont
  {Nicklow}}]{lattice-constant}%
  \BibitemOpen
  \bibfield  {author} {\bibinfo {author} {\bibfnamefont {N.}~\bibnamefont
  {Wakabayashi}}, \bibinfo {author} {\bibfnamefont {H.~G.}\ \bibnamefont
  {Smith}}, \ and\ \bibinfo {author} {\bibfnamefont {R.~M.}\ \bibnamefont
  {Nicklow}},\ }\href {\doibase 10.1103/PhysRevB.12.659} {\bibfield  {journal}
  {\bibinfo  {journal} {Phys. Rev. B}\ }\textbf {\bibinfo {volume} {12}},\
  \bibinfo {pages} {659} (\bibinfo {year} {1975})}\BibitemShut {NoStop}%
\bibitem [{\citenamefont {Lee}\ \emph {et~al.}(2010)\citenamefont {Lee},
  \citenamefont {Yan}, \citenamefont {Brus}, \citenamefont {Heinz},
  \citenamefont {Hone},\ and\ \citenamefont {Ryu}}]{lee2010anomalous}%
  \BibitemOpen
  \bibfield  {author} {\bibinfo {author} {\bibfnamefont {C.}~\bibnamefont
  {Lee}}, \bibinfo {author} {\bibfnamefont {H.}~\bibnamefont {Yan}}, \bibinfo
  {author} {\bibfnamefont {L.~E.}\ \bibnamefont {Brus}}, \bibinfo {author}
  {\bibfnamefont {T.~F.}\ \bibnamefont {Heinz}}, \bibinfo {author}
  {\bibfnamefont {J.}~\bibnamefont {Hone}}, \ and\ \bibinfo {author}
  {\bibfnamefont {S.}~\bibnamefont {Ryu}},\ }\href@noop {} {\bibfield
  {journal} {\bibinfo  {journal} {ACS Nano}\ }\textbf {\bibinfo {volume} {4}},\
  \bibinfo {pages} {2695} (\bibinfo {year} {2010})}\BibitemShut {NoStop}%
\bibitem [{\citenamefont {Tongay}\ \emph {et~al.}(2013)\citenamefont {Tongay},
  \citenamefont {Suh}, \citenamefont {Ataca}, \citenamefont {Fan},
  \citenamefont {Luce}, \citenamefont {Kang}, \citenamefont {Liu},
  \citenamefont {Ko}, \citenamefont {Raghunathanan}, \citenamefont {Zhou} \emph
  {et~al.}}]{tongay2013defects}%
  \BibitemOpen
  \bibfield  {author} {\bibinfo {author} {\bibfnamefont {S.}~\bibnamefont
  {Tongay}}, \bibinfo {author} {\bibfnamefont {J.}~\bibnamefont {Suh}},
  \bibinfo {author} {\bibfnamefont {C.}~\bibnamefont {Ataca}}, \bibinfo
  {author} {\bibfnamefont {W.}~\bibnamefont {Fan}}, \bibinfo {author}
  {\bibfnamefont {A.}~\bibnamefont {Luce}}, \bibinfo {author} {\bibfnamefont
  {J.~S.}\ \bibnamefont {Kang}}, \bibinfo {author} {\bibfnamefont
  {J.}~\bibnamefont {Liu}}, \bibinfo {author} {\bibfnamefont {C.}~\bibnamefont
  {Ko}}, \bibinfo {author} {\bibfnamefont {R.}~\bibnamefont {Raghunathanan}},
  \bibinfo {author} {\bibfnamefont {J.}~\bibnamefont {Zhou}},  \emph {et~al.},\
  }\href@noop {} {\bibfield  {journal} {\bibinfo  {journal} {Scientific
  Reports}\ }\textbf {\bibinfo {volume} {3}},\ \bibinfo {pages} {2657}
  (\bibinfo {year} {2013})}\BibitemShut {NoStop}%
\bibitem [{\citenamefont {Zhao}\ \emph {et~al.}(2013)\citenamefont {Zhao},
  \citenamefont {Ghorannevis}, \citenamefont {Amara}, \citenamefont {Pang},
  \citenamefont {Toh}, \citenamefont {Zhang}, \citenamefont {Kloc},
  \citenamefont {Tan},\ and\ \citenamefont {Eda}}]{zhao2013lattice}%
  \BibitemOpen
  \bibfield  {author} {\bibinfo {author} {\bibfnamefont {W.}~\bibnamefont
  {Zhao}}, \bibinfo {author} {\bibfnamefont {Z.}~\bibnamefont {Ghorannevis}},
  \bibinfo {author} {\bibfnamefont {K.~K.}\ \bibnamefont {Amara}}, \bibinfo
  {author} {\bibfnamefont {J.~R.}\ \bibnamefont {Pang}}, \bibinfo {author}
  {\bibfnamefont {M.}~\bibnamefont {Toh}}, \bibinfo {author} {\bibfnamefont
  {X.}~\bibnamefont {Zhang}}, \bibinfo {author} {\bibfnamefont
  {C.}~\bibnamefont {Kloc}}, \bibinfo {author} {\bibfnamefont {P.~H.}\
  \bibnamefont {Tan}}, \ and\ \bibinfo {author} {\bibfnamefont
  {G.}~\bibnamefont {Eda}},\ }\href@noop {} {\bibfield  {journal} {\bibinfo
  {journal} {Nanoscale}\ }\textbf {\bibinfo {volume} {5}},\ \bibinfo {pages}
  {9677} (\bibinfo {year} {2013})}\BibitemShut {NoStop}%
\bibitem [{\citenamefont {He}\ \emph {et~al.}(2014)\citenamefont {He},
  \citenamefont {Kumar}, \citenamefont {Zhao}, \citenamefont {Wang},
  \citenamefont {Mak}, \citenamefont {Zhao},\ and\ \citenamefont
  {Shan}}]{WSe2-he}%
  \BibitemOpen
  \bibfield  {author} {\bibinfo {author} {\bibfnamefont {K.}~\bibnamefont
  {He}}, \bibinfo {author} {\bibfnamefont {N.}~\bibnamefont {Kumar}}, \bibinfo
  {author} {\bibfnamefont {L.}~\bibnamefont {Zhao}}, \bibinfo {author}
  {\bibfnamefont {Z.}~\bibnamefont {Wang}}, \bibinfo {author} {\bibfnamefont
  {K.~F.}\ \bibnamefont {Mak}}, \bibinfo {author} {\bibfnamefont
  {H.}~\bibnamefont {Zhao}}, \ and\ \bibinfo {author} {\bibfnamefont
  {J.}~\bibnamefont {Shan}},\ }\href {\doibase 10.1103/PhysRevLett.113.026803}
  {\bibfield  {journal} {\bibinfo  {journal} {Phys. Rev. Lett.}\ }\textbf
  {\bibinfo {volume} {113}},\ \bibinfo {pages} {026803} (\bibinfo {year}
  {2014})}\BibitemShut {NoStop}%
\bibitem [{\citenamefont {Chellappan}\ \emph {et~al.}(2018)\citenamefont
  {Chellappan}, \citenamefont {Pang}, \citenamefont {Sarkar}, \citenamefont
  {Ooi},\ and\ \citenamefont {Goh}}]{chellappan2018effect}%
  \BibitemOpen
  \bibfield  {author} {\bibinfo {author} {\bibfnamefont {V.}~\bibnamefont
  {Chellappan}}, \bibinfo {author} {\bibfnamefont {A.~L.~C.}\ \bibnamefont
  {Pang}}, \bibinfo {author} {\bibfnamefont {S.}~\bibnamefont {Sarkar}},
  \bibinfo {author} {\bibfnamefont {Z.~E.}\ \bibnamefont {Ooi}}, \ and\
  \bibinfo {author} {\bibfnamefont {K.~E.~J.}\ \bibnamefont {Goh}},\
  }\href@noop {} {\bibfield  {journal} {\bibinfo  {journal} {Electronic
  Materials Letters}\ }\textbf {\bibinfo {volume} {14}},\ \bibinfo {pages}
  {766} (\bibinfo {year} {2018})}\BibitemShut {NoStop}%
\bibitem [{\citenamefont {Jones}\ \emph {et~al.}(2013)\citenamefont {Jones},
  \citenamefont {Yu}, \citenamefont {Ghimire}, \citenamefont {Wu},
  \citenamefont {Aivazian}, \citenamefont {Ross}, \citenamefont {Zhao},
  \citenamefont {Yan}, \citenamefont {Mandrus}, \citenamefont {Xiao} \emph
  {et~al.}}]{jones2013optical}%
  \BibitemOpen
  \bibfield  {author} {\bibinfo {author} {\bibfnamefont {A.~M.}\ \bibnamefont
  {Jones}}, \bibinfo {author} {\bibfnamefont {H.}~\bibnamefont {Yu}}, \bibinfo
  {author} {\bibfnamefont {N.~J.}\ \bibnamefont {Ghimire}}, \bibinfo {author}
  {\bibfnamefont {S.}~\bibnamefont {Wu}}, \bibinfo {author} {\bibfnamefont
  {G.}~\bibnamefont {Aivazian}}, \bibinfo {author} {\bibfnamefont {J.~S.}\
  \bibnamefont {Ross}}, \bibinfo {author} {\bibfnamefont {B.}~\bibnamefont
  {Zhao}}, \bibinfo {author} {\bibfnamefont {J.}~\bibnamefont {Yan}}, \bibinfo
  {author} {\bibfnamefont {D.~G.}\ \bibnamefont {Mandrus}}, \bibinfo {author}
  {\bibfnamefont {D.}~\bibnamefont {Xiao}},  \emph {et~al.},\ }\href@noop {}
  {\bibfield  {journal} {\bibinfo  {journal} {Nature Nanotechnology}\ }\textbf
  {\bibinfo {volume} {8}},\ \bibinfo {pages} {634} (\bibinfo {year}
  {2013})}\BibitemShut {NoStop}%
\bibitem [{\citenamefont {Schrott}\ \emph {et~al.}(2001)\citenamefont
  {Schrott}, \citenamefont {Misewich}, \citenamefont {Copel}, \citenamefont
  {Abraham},\ and\ \citenamefont {Zhang}}]{schrott2001site}%
  \BibitemOpen
  \bibfield  {author} {\bibinfo {author} {\bibfnamefont {A.}~\bibnamefont
  {Schrott}}, \bibinfo {author} {\bibfnamefont {J.}~\bibnamefont {Misewich}},
  \bibinfo {author} {\bibfnamefont {M.}~\bibnamefont {Copel}}, \bibinfo
  {author} {\bibfnamefont {D.}~\bibnamefont {Abraham}}, \ and\ \bibinfo
  {author} {\bibfnamefont {Y.}~\bibnamefont {Zhang}},\ }\href@noop {}
  {\bibfield  {journal} {\bibinfo  {journal} {Applied Physics Letters}\
  }\textbf {\bibinfo {volume} {79}},\ \bibinfo {pages} {1786} (\bibinfo {year}
  {2001})}\BibitemShut {NoStop}%
\bibitem [{\citenamefont {Barbone}\ \emph {et~al.}(2018)\citenamefont
  {Barbone}, \citenamefont {Montblanch}, \citenamefont {Kara}, \citenamefont
  {Palacios-Berraquero}, \citenamefont {Cadore}, \citenamefont {De~Fazio},
  \citenamefont {Pingault}, \citenamefont {Mostaani}, \citenamefont {Li},
  \citenamefont {Chen} \emph {et~al.}}]{barbone2018complexes}%
  \BibitemOpen
  \bibfield  {author} {\bibinfo {author} {\bibfnamefont {M.}~\bibnamefont
  {Barbone}}, \bibinfo {author} {\bibfnamefont {A.~R.-P.}\ \bibnamefont
  {Montblanch}}, \bibinfo {author} {\bibfnamefont {D.~M.}\ \bibnamefont
  {Kara}}, \bibinfo {author} {\bibfnamefont {C.}~\bibnamefont
  {Palacios-Berraquero}}, \bibinfo {author} {\bibfnamefont {A.~R.}\
  \bibnamefont {Cadore}}, \bibinfo {author} {\bibfnamefont {D.}~\bibnamefont
  {De~Fazio}}, \bibinfo {author} {\bibfnamefont {B.}~\bibnamefont {Pingault}},
  \bibinfo {author} {\bibfnamefont {E.}~\bibnamefont {Mostaani}}, \bibinfo
  {author} {\bibfnamefont {H.}~\bibnamefont {Li}}, \bibinfo {author}
  {\bibfnamefont {B.}~\bibnamefont {Chen}},  \emph {et~al.},\ }\href@noop {}
  {\bibfield  {journal} {\bibinfo  {journal} {Nature Comm.}\ }\textbf {\bibinfo
  {volume} {9}},\ \bibinfo {pages} {3721} (\bibinfo {year} {2018})}\BibitemShut
  {NoStop}%
\end{thebibliography}%

\begin{center}
{\bf \large  Supplemental Material} 
\end{center}

{\bf Sample Growth} ---
The growth of MoS$_2$ was carried out using 4 inch quartz tube inside a single-zone tube furnace (Thermcraft). Sulfur powder (Sigma-Aldrich, 99.99\% pure) was kept outside the hot zone and MoO$_3$ powder (Sigma-Aldrich, 99.999\% pure) and high purity STO substrates of different crystal orientations (CrysTec GmbH) were placed at the center of the hot-zone. Atmospheric pressure CVD growth is performed using high purity nitrogen gas as the carrier gas. The growth temperature was at 700$^\circ$ C, and we followed the method by van der Zande et al. \cite{NatMat2013growth}. 

{\bf Optical Spectroscopy Methods} --- 
The Raman and PL spectroscopy were carried out using JY Horiba LabRAM HR Evolution Raman Spectrometer in the backscattering geometry. The excitation radiation was 514.5 nm from a Lexel SHG 95 Argon ion laser. To avoid any thermal effects, the power of the laser incident on our sample was less than 500 $\mu$W for the above measurements. A long working distance 50X objective was used for the above measurements and the spot size of the laser beam was focused to 1-2 $\mu$m. A helium closed cycle refrigerator cooled cryogenic stage with optically transparent window were used to collect PL spectra from 300 K to 10 K. A computer controlled motorized XY stage was used in order to record the PL mapping of the sample under study.

{\bf Polaronic trion binding energy fitting} ---
The bare trion binding energy can be estimated perturbatively \cite{efimkin2017}, but
we assume the phenomenological formula \cite{binding-energy-fit-2014}
$E_T(\epsilon_0)=E_T(1)/\epsilon_0^{\beta_T}$,
where $\beta_T=0.085$ is a constant, $E_T(1)= 42$ meV is $E_T$ at $\epsilon_0=1$. 
The polaronic contribution is calculated using the theory introduced in the main text and
material constants discussed below.

Static dielectric screening on the MoS$_2$/STO interface is largely dominated by STO,
especially at low temperatures, when the relative dielectric permittivity exceeds $10^4$ \cite{STOdielectric1962}. This is governed by the Curie-Weiss law modified by Barrett \cite{STO-dielectric-formula1952}, and we set accordingly
$$
\epsilon_0= \frac{T_2}{\frac{T_1}{2}\coth\left(\frac{T_1}{2T}\right) - T_C},
$$
where $T_2=9\cdot 10^4$ K, $T_1= 84$ K, and Curie temperature $T_C= 38$ K.
In contrast, the high-frequency screening is dominated by the 2D semiconductor \cite{2DMoS2dielectric2018} with the in-plane dielectric permittivity $\epsilon_\infty = 15.1$, substantially higher than that for STO \cite{STO-high-freq-diel-1997}.
The RO phonon frequency is dominated by the $A_{1g}$ mode with the temperature dependence given by
\cite{STO-soft-phonon-temprature2001,STO-soft-phonon-temprature1968}
$\omega= \omega_0 \left(1- T/T_a\right)^\frac{1}{3}$, where $T \leq T_a$,  $\hbar\omega_0 = 7$ meV, and the mode activation temperature is $T_a=132$ K.

We emphasize that our activation temperature $T_a=132$ K is different from the STO phase transition temperature around $105$ K. This is because the interfacial nature of the effect considered. The trions excited in 2D MoS$_2$ interact with the rotational phonons located on or very near to the STO surface, not with the bulk ones. As it is shown in Ref. \cite{STOdepth2011}, the phase transition starts in the near-surface regions at temperatures much higher than the bulk transition temperature of about $105$ K. In fact, it can be as high as $155$ K for some surface domains. The higher surface transition temperature is probably responsible for the higher activation temperature of the rotational $A_{1g}$ phonon mode found by means of Raman spectroscopy  \cite{STO-soft-phonon-temprature2001}.

The difference between the surface and bulk critical temperatures
can be seen in our measurements: The trionic binding energy stops decreasing with temperature near $150$ K, but the trend is not obvious until the temperature drops below $135$ K. At $T<105$ K the transition is complete, and the temperature dependence of the binding energy is determined by that of the soft phonon frequency and static dielectric constant. Our model does not distinguish the domains by their preferred transition temperature, and this is the reason why the theoretical binding energy is such a sharp function of $T$.  We could try to modify the theory to account for different domains switching at different temperatures giving some broadening; however, we choose instead to keep the theory without any fitting parameters.

The remaining  material constants for MoS$_2$ represent 
effective masses $m_e=0.46 m_0$, $m_h = 0.56 m_0$  
\cite{PRB2013berkelbach,lu2014van,eff-mass2016} 
($m_0$ is the free electron mass), and
the lattice constant $a=3.15\,\mathrm{\mathring{A}}$ \cite{lattice-constant}.
As far as WSe$_2$ is concerned, the high-frequency dielectric permittivity 
remains the same as for MoS$_2$ \cite{2DMoS2dielectric2018},
but the effective masses are somewhat lower,
namely, $m_e=0.33 m_0$, $m_h = 0.43 m_0$ \cite{PRB2013berkelbach,eff-mass2016}.

Since there are no adjustable parameters, the minor quantitative disagreement should not take away from our main conclusion about the dependence of binding energy on substrate orientation, for which the theory and experiment are well matched.

\begin{figure}
\renewcommand{\thefigure}{{\bf S1}}
 \includegraphics[width=\columnwidth]{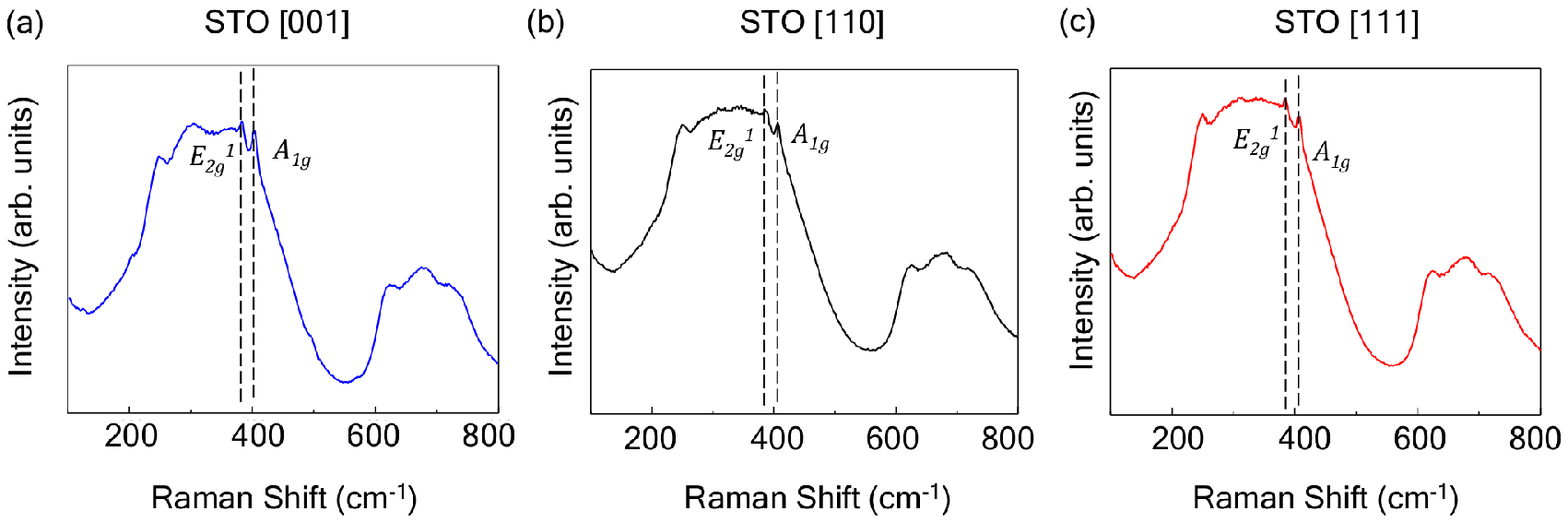}
 \caption{Raman spectra of monolayer MoS$_2$ grown on (a) STO [001] (b) STO [110] and (c) STO [111]. The black dashed lines indicate the characteristic Raman modes for 2H-MoS$_2$ with the strong second order vibrational modes of STO \cite{STO-soft-phonon-temprature2001} in the background. The separation between the in-plane (E$_{2g}$) and out-of-plane (A$_{1g}$) vibrational modes of MoS$_2$ is $\sim 20$ cm$^{-1}$, which indicates that our sample is a monolayer \cite{lee2010anomalous}.}
 \label{figS1}
\renewcommand{\thefigure}{{\bf S2}}
 \includegraphics[width=\columnwidth]{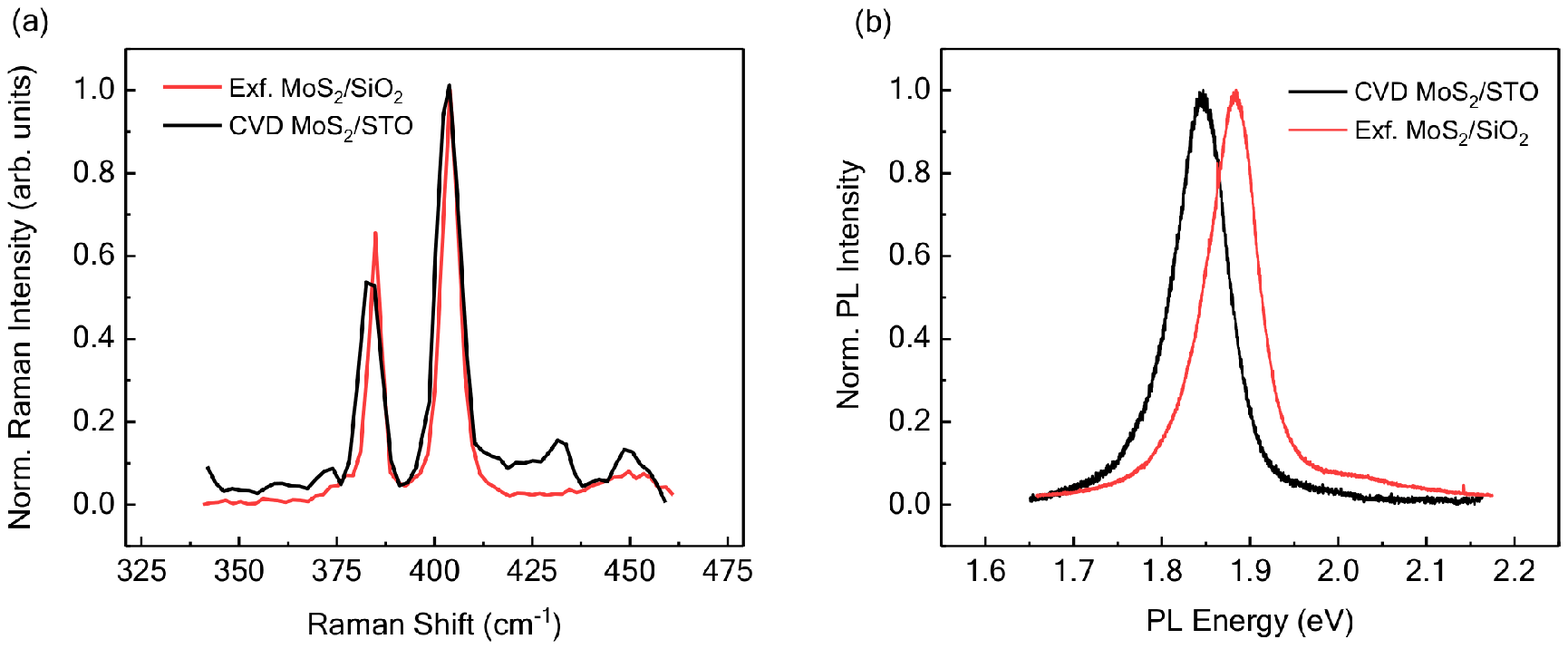}
 \caption{Comparison of (a) Raman and (b) PL spectra of an exfoliated monolayer MoS$_2$ (in our lab) with MoS$_2$/STO. In the Raman spectra, both the characteristic in-plane E$_{2g}$ and out-of-plane A$_{1g}$­ modes in MoS$_2$/STO (after background correction of STO Raman spectra) and MoS$_2$/SiO$_2$ show a similar linewidth and intensity ratio that suggests that the sample is of good quality and there is no substrate induced charge doping in MoS$_2$/STO. The PL spectra too shows identical spectral shape and linewidth. The PL peak in MoS$_2$/STO is redshifted as compared to the PL emission on MoS$_2$/SiO$_2$ due to dielectric screening \cite{binding-energy-fit-2014} from the substrate.}
 \label{figS2}
\end{figure}

\begin{center}
{\bf \large Section S1: Sample Characterization}
\end{center}
See Figures \ref{figS1} and \ref{figS2}.

\renewcommand{\thefigure}{{\bf S3}}
\begin{figure}[b]
 \includegraphics[width=\columnwidth]{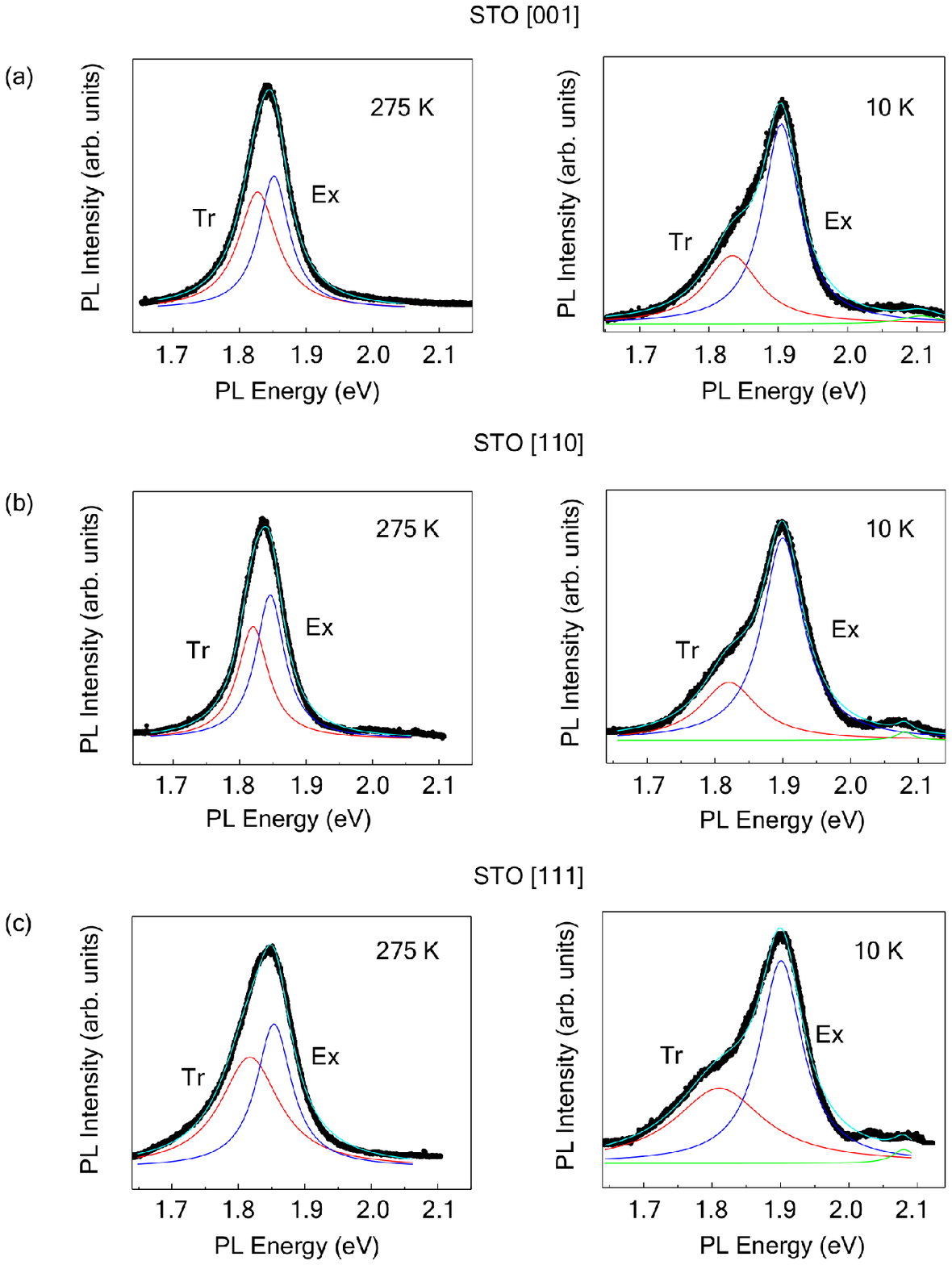}
 \caption{PL spectra at 275 K (left) and 10 K (right) fitted with 3 different Lorentzian peaks for MoS$_2$ on (a) STO [001] (b) STO [110] and (c) STO [111]. The cumulative curve fit shows a close match with the experimental data. Red curve: A$^-$ trion; blue curve: A exciton; green curve: B exciton; cyan curve: cumulative fit.}
 \label{figS3}
\end{figure}

\begin{center}
{\bf \large Section S2: Lorentz Fitting of PL Spectra}
\end{center}

{\bf Fitting Principle} --- We have used the standard curve fitting software to fit our MoS$_2$ PL spectra. As the PL of MoS$_2$ cannot be fitted using a single Lorentzian curve due to spectral asymmetry we use 3 Lorentzian curves (standard practice in the community) corresponding to the A exciton, A$^-$ trion, and B exciton.
The Lorentzian curves are initialized and then allowed to adjust through multiple iterations to obtain the best cumulative fit that closely matches the experimental raw data. For all our fits, the coefficient of determination in the regression model COD (R2) $\sim 0.99$ providing a satisfactory fit, as shown in Figure \ref{figS3}.

{\bf Determination of Peak Position} --- The peak position of each individual quasiparticle has been determined from the peak center of their corresponding Lorentz spectrum. 

{\bf Determination of Peak Width} --- The linewidth of each individual quasiparticle has been determined from the FWHM of their corresponding Lorentz spectrum.

{\bf Table S1} --- Varshni fit \cite{VARSHNI1967149}, $E=E_0 - AT^2/(T+B)$, parameters for the A exciton (upper) and A$^-$ trion (lower) energy in MoS$_2$ grown on different substrate orientations.
\begin{center}
\begin{tabular}{c|c|c|c}
Sample               &  $E_0$ (eV) & $A$ (eV/K$^2$)      & $B$ (K) \\
\hline
MoS$_2$/STO [001]  & $1.906$    & $4.55\cdot 10^{-4}$ & $480$ \\
MoS$_2$/STO [011]  & $1.902$    & $5.45\cdot 10^{-4}$ & $569$ \\
MoS$_2$/STO [111]  & $1.906$    & $4.1\cdot 10^{-4}$  & $310$ \\
\hline
MoS$_2$/STO [001]  & $1.868$    & $2.55\cdot 10^{-4}$ & $482$ \\
MoS$_2$/STO [011]  & $1.875$    & $3.93\cdot 10^{-4}$ & $360$ \\
MoS$_2$/STO [111]  & $1.873$    & $4.2\cdot 10^{-4}$  & $307$\\
\hline
\end{tabular}
\end{center}

\begin{center}
{\bf \large Section S3: Excitation Power Dependent PL Spectra}
\end{center}
See Figure \ref{figS4} below.

\renewcommand{\thefigure}{{\bf S4}}
\begin{figure}
 \includegraphics[width=0.7\columnwidth]{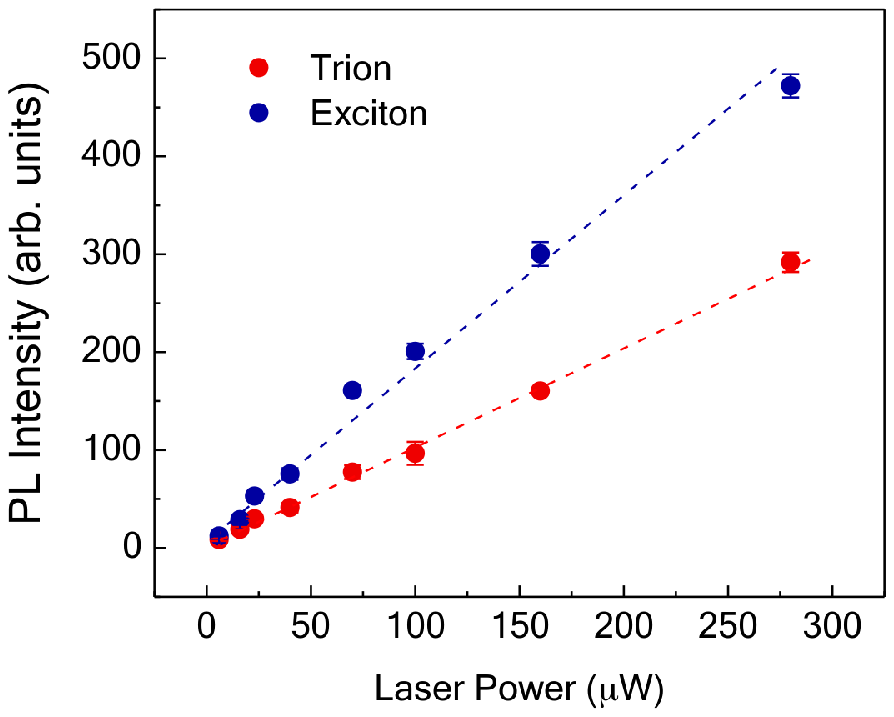}
 \caption{PL intensity of the exciton and trion in MoS$_2$/STO [001] as a function of incident laser excitation power at 10 K. Both the exciton and trion peaks have a linear power dependence, as it has been reported previously, in contrast to the defect emission, which is expected to saturate at higher powers when the defects are fully populated with excitons \cite{tongay2013defects}. }
 \label{figS4}
\end{figure}

\begin{center}
{\bf \large Section S4: Identification of Trions by Electrostatic Gating }
\end{center}

\renewcommand{\thefigure}{{\bf S5}}
\begin{figure}
 \includegraphics[width=\columnwidth]{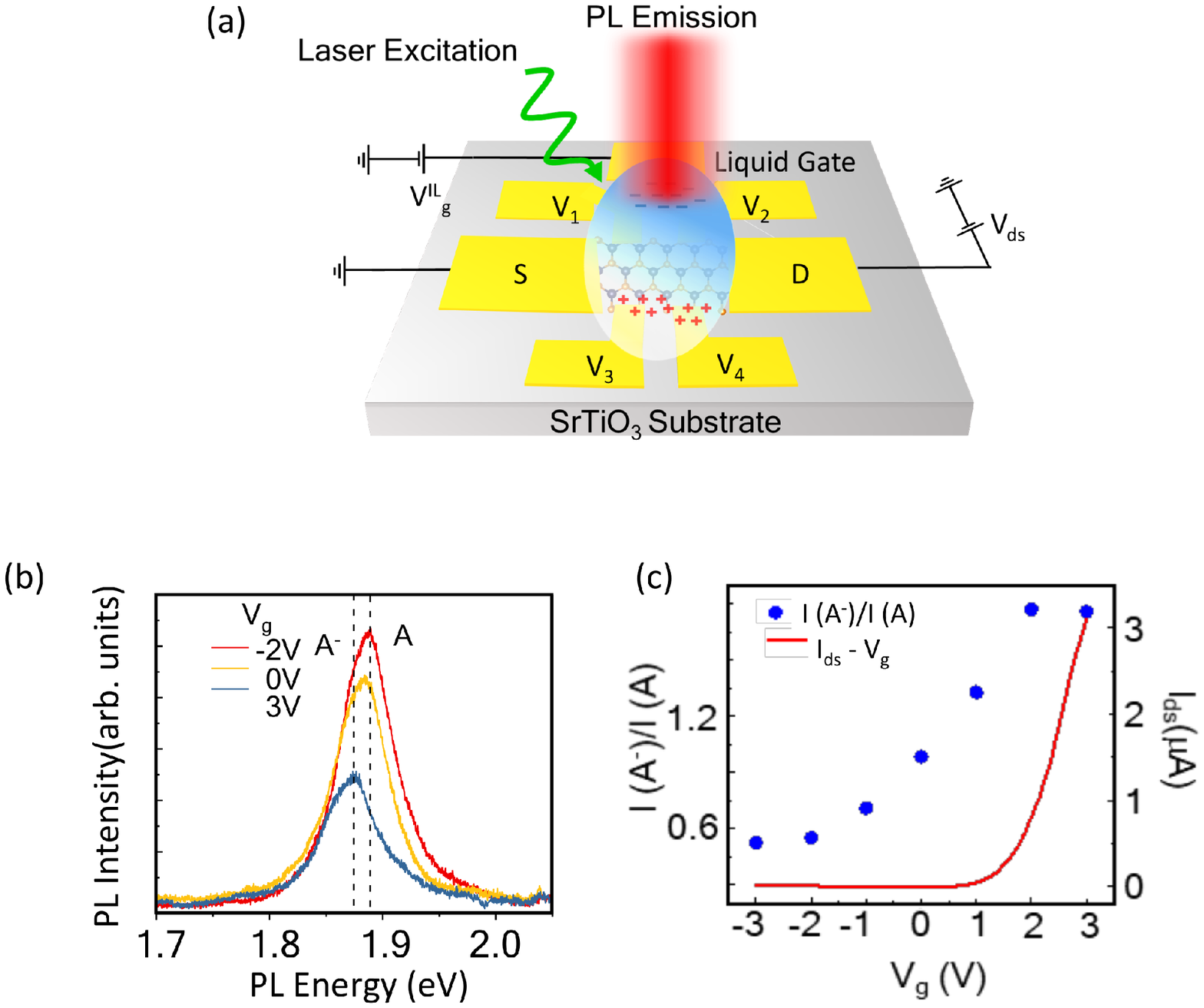}
 \caption{(a) Schematics of our device for measuring PL emission as a function of electrostatic gate voltage. (b) PL spectra as a function of top gate voltage. (c) Intensity ratio of the trion to exciton component (blue dots) of the PL spectra as a function of top gate voltage correlated with the $I_{ds}$--$V_g$ characteristics (red curve).}
 \label{figS5}
\end{figure}

We have performed electrostatic gate dependent measurement of the PL intensity to confirm the identity of the quasiparticles. As shown in Figure \ref{figS5}(c), when we move towards the electron rich $n$--regime (evident from the $I_{ds}$--$V_g$ characteristics), the intensity ratio of the trion to exciton peaks in the PL spectra increases, a phenomena generally associated to the spectral weight transfer from the neutral exciton to the charged trion \cite{mak2013trions}. This provides definitive evidence of the charge nature of the peaks and confirms that the measured spectra are due to trions rather than defects.

\begin{figure}[b]
\renewcommand{\thefigure}{{\bf S6}}
 \includegraphics[width=\columnwidth]{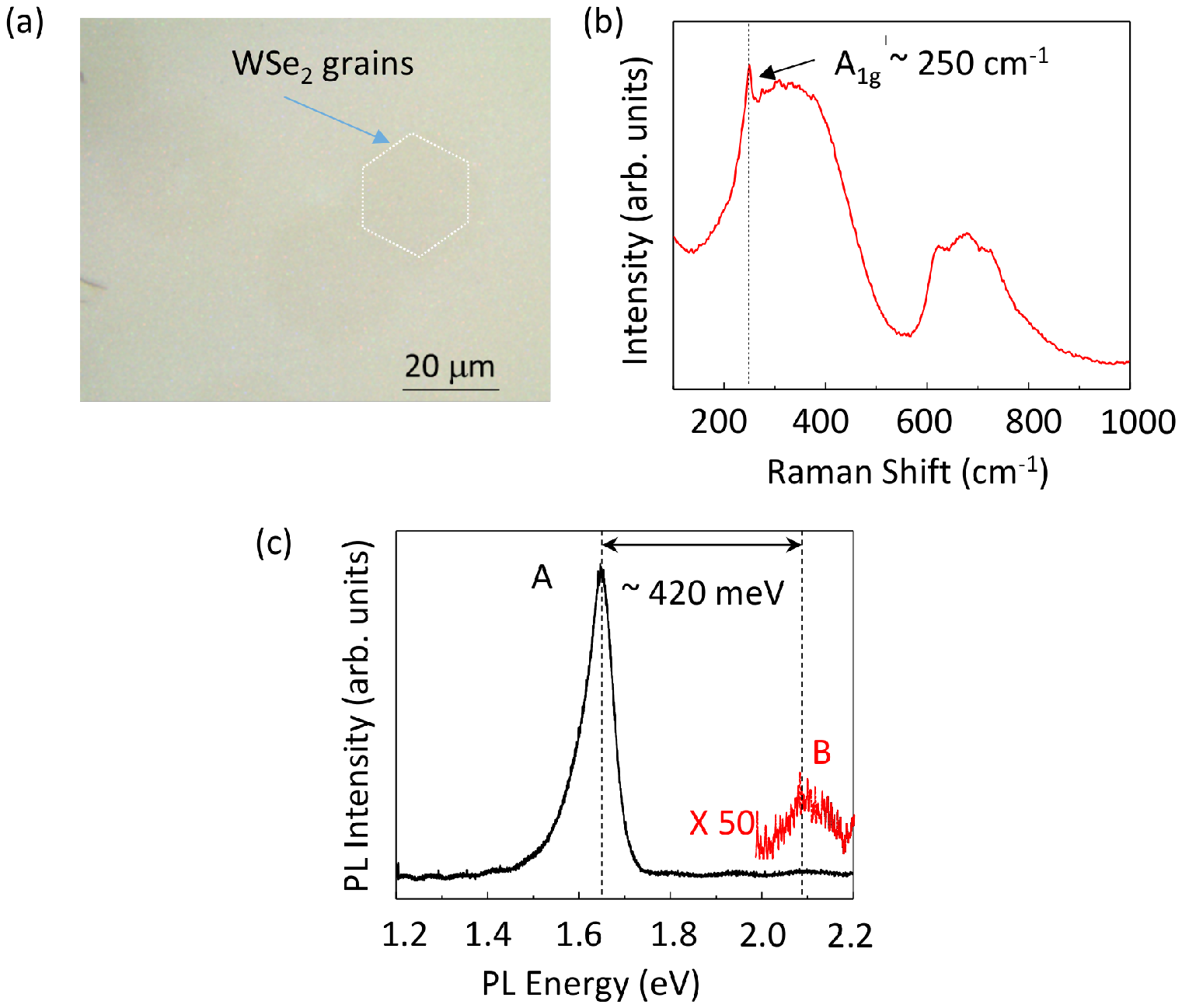}
 \caption{(a) Optical image of WSe$_2$/STO. The white hexagon is a guide to the eye to help locate the WSe$_2$ crystals that have poor optical contrast due to the transparent STO substrate. (b) Raman spectra of WSe$_2$/STO. The characteristic A$_{1g}$ mode \cite{zhao2013lattice} is observed at $\sim 250$ cm$^{-1}$ with a wide background that is associated with the STO Raman spectra. (c) PL spectra of monolayer WSe$_2$/STO shows a direct bandgap emission at $1.65$ eV and a weak B exciton $\sim 2.1$ eV consistent with previous reports from high quality samples \cite{WSe2-he,chellappan2018effect}.}
 \label{figS6}
\renewcommand{\thefigure}{{\bf S7}}
 \includegraphics[width=\columnwidth]{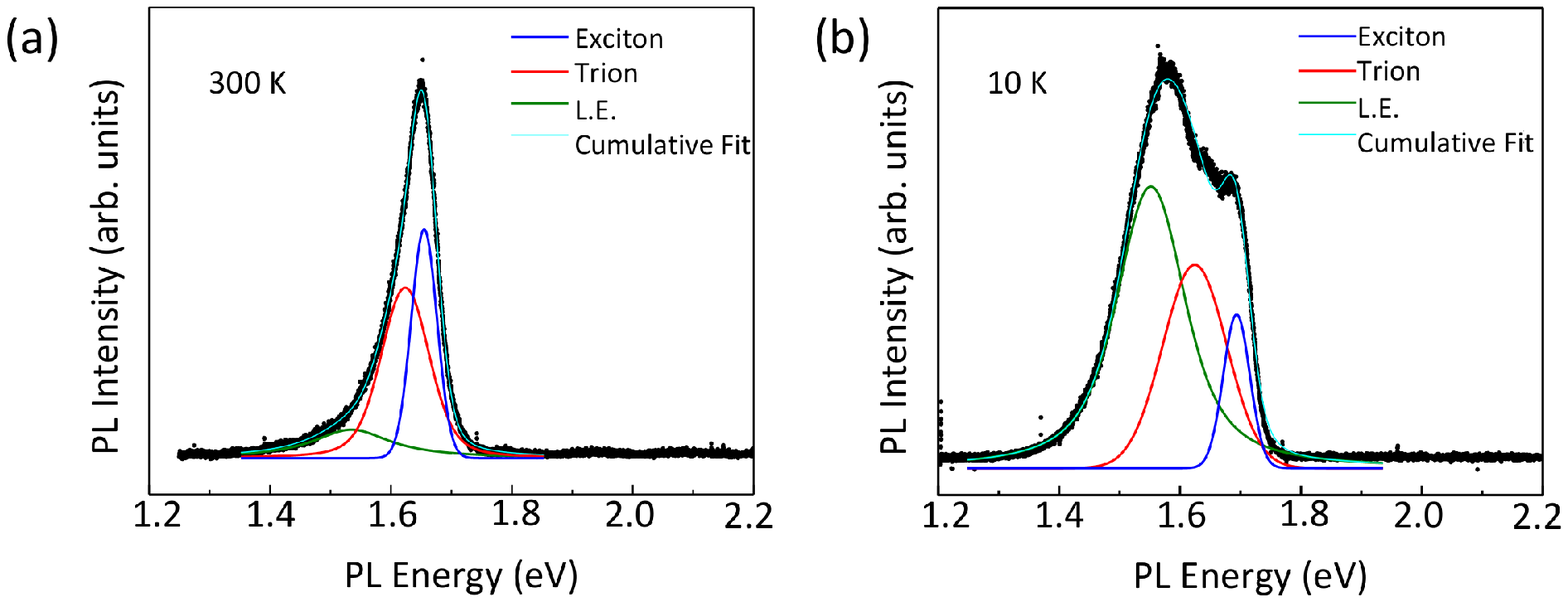}
 \caption{PL Spectra of WSe$_2$/STO at (a) 300 K and (b) 10 K fitted with three peaks (consistent with previous reports) that represent the neutral exciton (blue), charged trion (red), and localized exciton (L.E.). The cyan curve is the cumulative fit. One can observe spectral broadening and enhancement of the separation of the excitonic and trionic emission energies, that translates to a higher trion binding energy (TBE). This is in clear contrast to several reports \cite{chellappan2018effect,jones2013optical} on WSe$_2$/SiO$_2$ that show sharp excitonic resonances and absence of TBE enhancement of low temperature, suggesting polaronic phenomena at the WSe$_2$/STO interface. }
 \label{figS7}
\end{figure}

\begin{center}
{\bf \large  Section S5: Investigating WSe$_2$/STO}
\end{center}
See Figures \ref{figS6} and \ref{figS7}.

\end{document}